\documentclass[twocolumn]{aastex63}

\usepackage[fleqn]{amsmath}
\usepackage[hyphenbreaks]{breakurl}
\usepackage{float}

\makeatletter
\newcommand*{\rom}[1]{\expandafter\@slowromancap\romannumeral #1@}
\makeatother

\received{Jul 2, 2020}
\revised{Aug 23, 2021}
\accepted{Aug 25, 2021}
\submitjournal{ApJ}

\shorttitle{Variable stars in M15}
\shortauthors{Bhardwaj A. et al.}

\begin{document}
\title{Optical and near-infrared pulsation properties of RR Lyrae and Population II Cepheid variables in the Messier 15 globular cluster}

\correspondingauthor{Anupam Bhardwaj}
\email{anupam.bhardwajj@gmail.com; abhardwaj@kasi.re.kr}
\author[0000-0001-6147-3360]{Anupam Bhardwaj}\thanks{IAU Gruber Foundation Fellow}\thanks{EACOA Fellow}
\affil{Korea Astronomy and Space Science Institute, Daedeokdae-ro 776, Yuseong-gu, Daejeon 34055, Republic of Korea}
\author[0000-0002-6577-2787]{Marina Rejkuba}
\affiliation{European Southern Observatory, Karl-Schwarzschild-Stra\ss e 2, 85748, Garching, Germany}
\author[0000-0003-4520-1044]{G. C. Sloan}
\affiliation{Space Telescope Science Institute, 3700 San Martin Drive, Baltimore, MD 21218, USA}
\affiliation{Department of Physics and Astronomy, University of North Carolina, Chapel Hill, NC 27599-3255, USA}
\author[0000-0002-1330-2927]{Marcella Marconi}
\affil{INAF-Osservatorio astronomico di Capodimonte, Via Moiariello 16, 80131 Napoli, Italy}
\author[0000-0001-9842-639X]{Soung-Chul Yang}
\affiliation{Korea Astronomy and Space Science Institute, Daedeokdae-ro 776, Yuseong-gu, Daejeon 34055, Republic of Korea}

\begin{abstract} 
Messier 15 (NGC 7078) is an old and metal-poor post core-collapse globular cluster which hosts a rich population of variable stars. We report new optical ($gi$) and near-infrared (NIR, $JK_s$) multi-epoch observations for 129 RR Lyrae, 4 Population II Cepheids (3 BL Herculis, 1 W Virginis), and 1 anomalous Cepheid variable candidate in M15 obtained using the MegaCam and the WIRCam instruments on the 3.6-m Canada--France--Hawaii Telescope. Multi-band data are used to improve the periods and classification of variable stars, and determine accurate mean magnitudes and pulsational amplitudes from the light curves fitted with optical and NIR templates. We derive optical and NIR period--luminosity relations for RR Lyrae stars which are best constrained in the $K_s$-band, $m_{K_s} = -2.333~(0.054) \log P + 13.948~(0.015)$ with a scatter of only $0.037$ mag. Theoretical and empirical calibrations of RR Lyrae period--luminosity--metallicity relations are used to derive a true distance modulus to M15: $15.196~\pm~0.026$~(statistical)~$\pm~ 0.039$~(systematic) mag. Our precise distance moduli based on RR Lyrae stars and Population II Cepheid variables are mutually consistent and agree with recent distance measurements in the literature based on {\it Gaia} parallaxes and other independent methods.\\
\end{abstract} 

\section{Introduction}
RR Lyrae stars and Population II Cepheids are radially pulsating variable stars that are excellent distance indicators and useful tracers of old and metal-poor stellar populations. RR Lyrae variables are low-mass stars (0.5--0.8$M_\odot$) that are located within the intersection of the horizontal branch and the instability strip in the Hertzsprung--Russell diagram. BL Herculis and W Virginis stars represent two subclasses of Population II or Type II Cepheids in the post-horizontal-branch evolutionary phase \citep[see the review by][]{bhardwaj2020}. Anomalous Cepheids are also metal-poor but relatively massive  stars (1--2$M_\odot$) and are typically brighter than the horizontal branch RR Lyrae stars having similar colors on the color--magnitude diagram \citep{fiorentino2012, groenewegen2017}. 

The radially pulsating stars obey a well-defined period--luminosity relation (PLR) with a smaller dispersion at near-infrared (NIR) wavelengths as compared to the optical bands due to less sensitivity to temperature variations within the instability strip \citep{catelan2004, marconi2015}, smaller amplitude variations, and less sensitivity to metallicity and extinction at longer wavelengths. RR Lyrae stars, in particular, follow a period--luminosity--metallicity (PLZ) relation with a significant dependence on metallicity at infrared wavelengths \citep{marconi2015, navarrete2017, braga2018, neeley2019, bhardwaj2021}.

Messier 15 (NGC 7078) is one of the most massive and luminous Galactic globular clusters (GCs) with $M=6.33\times10^5M_\odot$ and $V = 6.28\pm0.02$~mag \citep[][]{baumgardt2021}. It has a dense stellar core of radius $\sim 0.14\arcmin$ \citep{harris2010}. This old \citep[$\sim 12.5\pm 0.25$ Gyr;][]{vandenberg2016} and metal-poor \citep[{[Fe/H]}$\sim-2.3$~dex;][]{carretta2009, vandenberg2016} GC hosts nearly 200 candidate variable stars \citep{clement2001}. The horizontal branch of M15 spans a wide range of colors and is well-populated over the instability strip where more than 150 RR Lyrae stars are located \citep[see the catalog of][]{clement2001}. Other similar metal-poor GCs host a significantly smaller number of variables, most notably Messier 92 \citep[{[Fe/H]}$=-2.3$~dex, 17 RR Lyrae,][]{del2005}. 

In typical metal-poor GCs, the paucity of RR Lyrae variables, longer mean periods of fundamental-mode pulsators (RRab $\sim 0.65$ days), and a larger fraction of overtone-mode pulsators (RRc) can be explained by their evolution from the blue to the red side of the instability strip - characteristics of RR Lyrae stars in the Oosterhoff type II GCs \citep[OoII;][]{Oosterhoff1939, catelan2009, fabrizio2019, prudil2019}. Oosterhoff type I (OoI) GCs are more metal-rich ([Fe/H]$\gtrsim-1.6$~dex) and their RRab stars have shorter pulsation periods (mean $\sim 0.55$ days). Although the mean RRab period of M15 is consistent with other OoII type clusters, unlike those, the M15 RR Lyrae population indicates that their horizontal branch evolution may have initiated within the instability strip \citep{bingham1984,vandenberg2016}.

Optical photometry of M15 has now been carried out over the past century \citep[e.g.,][]{bailey1919, sandage1981, bingham1984, silbermann1995, corwin2008, hoffman2021}. Most of these optical studies are based on 1--2m class telescopes and focused on the identification, period determinations, and classifications of variable stars. \citet{siegel2015} investigated ultraviolet properties of the RR Lyrae population in M15 based on well-sampled light curves. In the NIR, photometry of M15 is predominantly limited to single-epoch observations mostly studying PLRs of RR Lyrae stars, color--magnitude diagrams and the properties of the red giants \citep[e.g.,][]{longmore1990, ferraro2000, ivanov2000, valenti2004, sollima2006, moneli2015}.  \citet{sollima2006} derived a $K_s$-band PLR for a sample of 52 RR Lyrae stars in M15 using photometry from \citet{longmore1990},  \citet{valenti2004}, and the Two Micron All Sky Survey \citep[2MASS;][]{skrutskie2006} and calibrated a PLZ$_{K_s}$ relation using RR Lyrae stars in 15 GCs. However, the variable star population of M15 has not been studied in detail using multi-epoch $JHK_s$ observations.
  
In this paper, we present optical ($gi$) and NIR ($JK_s$) multi-epoch photometry of RR Lyrae stars and Population II Cepheids for the first time in these filters and discuss their pulsation properties. This work is part of a larger project focusing on NIR multi-epoch observations of RR Lyrae in GCs, complementing studies of Messier 3 and Messier 53 \citep{bhardwaj2020a, bhardwaj2021} with a more metal-poor stellar population. Section~\ref{sec:data} describes the optical and NIR data, the data reduction, and the astrometric and photometric calibration. Section~\ref{sec:var} presents the variable stars and their optical and NIR light curves. The optical and NIR pulsation properties and the PLRs are discussed in Section~\ref{sec:puls} and a distance to M15 is derived in Section~\ref{sec:dis}. Section~\ref{sec:discuss} summarizes the main results of this work.

\section{Data and Photometry} \label{sec:data}

\subsection{Observations and data reduction}

This work is based on data in the science archive from the Canada--France--Hawaii Telescope (CFHT){\footnote{\url{http://www.cadc-ccda.hia-iha.nrc-cnrc.gc.ca/en/cfht/}}}. The optical and NIR images of M15 were taken between August 2011 and September 2013 under the program focused on pulsations of long-period variables and stellar deaths in globular clusters (PIs: G. Sloan \& D. Devost).  The short exposures (5s for MegaCam and 3s for WIRCam), originally adopted to avoid saturation of bright red giant branch variables, are optimal for our target horizontal-branch and post-horizontal-branch variables. 

\begin{figure*}
\epsscale{1.2}
\plotone{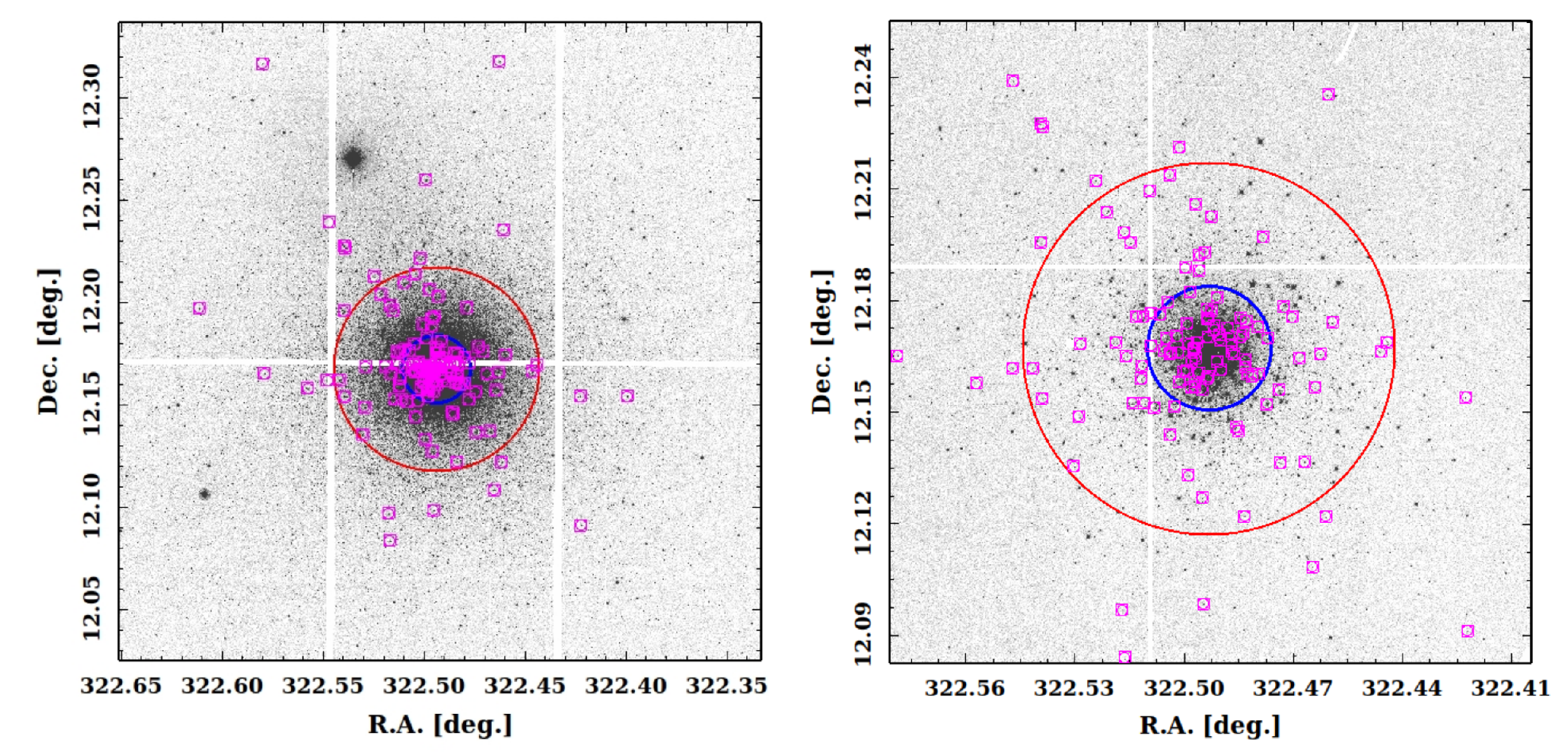}
\caption{{\it Left:} A mosaic $g$-band image created from 6 MegaCam CCDs  showing a region of $\sim19'\times 19'$ around the cluster center and small gaps of $13\arcsec$ between the CCDs. {\it Right:} A portion of the WIRCam image (detector $\#4,~\sim10'\times 10'$) displaying one of the individual $K_s$-band frames. Small (blue) and large (red) circles represent the half-light radius \citep[$r_h=1\arcmin$,][]{harris2010} and $3r_h$, respectively. The variable stars analyzed in this work are plotted with square (magenta) symbols. The variables that fall within the gaps between the MegaCam CCDs or in the bad pixels of WIRCam detector were recovered in other dithered frames.} 
\label{fig:fov}
\end{figure*}

\begin{deluxetable}{rrrrrrc}
\tablecaption{Log of Optical and NIR Observations. \label{tbl:log}}
\tabletypesize{\footnotesize}
\tablewidth{0pt}
\tablehead{\colhead{Date} & \colhead{MJD$_{\rm start}$} & \colhead{MJD$_{\rm end}$} & \colhead{AM} & \colhead{IQ (\arcsec)} & \colhead{$N_f$} & \colhead{ET (s)}}
\startdata
\multicolumn{7}{c}{$g$ band}\\
\hline
 2011-08-05&   778.476&   778.479&   1.02&   1.26&           6&           5\\
 2011-11-28&   893.193&   893.196&   1.06&   0.98&           6&           5\\
 2013-04-13&  1395.614&  1395.619&   1.65&   1.03&           9&           5\\
 2013-04-17&  1399.604&  1399.607&   1.64&   0.60&           7&           5\\
 2013-05-03&  1415.616&  1415.620&   1.20&   0.61&           7&           5\\
 2013-05-08&  1420.609&  1420.613&   1.17&   0.89&           7&           5\\
 2013-05-09&  1421.617&  1421.620&   1.13&   0.44&           7&           5\\
 2013-05-11&  1423.622&  1423.625&   1.10&   0.99&           7&           5\\
 2013-05-13&  1425.620&  1425.624&   1.09&   0.83&           7&           5\\
 2013-06-06&  1449.577&  1449.580&   1.04&   0.51&           7&           5\\
 2013-07-04&  1477.606&  1477.609&   1.09&   0.62&           7&           5\\
 2013-07-07&  1480.595&  1480.599&   1.08&   1.16&           7&           5\\
 2013-08-01&  1505.403&  1505.407&   1.08&   0.88&           7&           5\\
 2013-08-06&  1510.600&  1510.604&   1.63&   0.51&           7&           5\\
 2013-08-30&  1534.491&  1534.495&   1.26&   1.32&           7&           5\\
 2013-09-09&  1544.313&  1544.316&   1.05&   0.45&           7&           5\\
 \hline
 \multicolumn{7}{c}{$i$ band}\\
\hline
 2011-11-28&   893.188&   893.191&   1.05&   0.90&           6&           5\\
 2013-04-13&  1395.620&  1395.629&   1.55&   1.03&          10&           5\\
 2013-04-17&  1399.609&  1399.613&   1.58&   0.68&           7&           5\\
 2013-05-03&  1415.622&  1415.625&   1.18&   0.67&           7&           5\\
 2013-05-08&  1420.614&  1420.618&   1.15&   0.81&           7&           5\\
 2013-05-09&  1421.622&  1421.624&   1.12&   0.51&           4&           5\\
 2013-05-10&  1422.594&  1422.595&   1.22&   0.87&           3&           5\\
 2013-05-14&  1426.605&  1426.609&   1.12&   0.81&           7&           5\\
 2013-06-06&  1449.582&  1449.586&   1.03&   0.42&           7&           5\\
 2013-07-07&  1480.600&  1480.604&   1.10&   1.12&           7&           5\\
 2013-08-01&  1505.409&  1505.412&   1.07&   0.88&           7&           5\\
 2013-08-06&  1510.606&  1510.609&   1.70&   0.54&           7&           5\\
 2013-08-30&  1534.496&  1534.500&   1.29&   1.42&           7&           5\\
 2013-09-09&  1544.318&  1544.322&   1.04&   0.48&           7&           5\\
 \hline
 \multicolumn{7}{c}{$J$ band}\\
\hline
 2011-09-08&   812.245&   812.250&   1.32&   0.71&          18&           3\\
 2011-09-27&   831.235&   831.239&   1.12&   0.43&          17&           3\\
 2011-10-03&   837.227&   837.231&   1.10&   0.70&          17&           3\\
 2011-10-03&   837.417&   837.421&   1.38&   0.63&          17&           3\\
 2011-10-04&   838.198&   838.212&   1.20&   0.68&          10&           3\\
 2011-10-04&   838.218&   838.222&   1.12&   0.68&          17&           3\\
 2011-10-04&   838.223&   838.227&   1.10&   0.44&          16&           3\\
 2011-10-05&   839.223&   839.227&   1.09&   0.56&          18&           3\\
 2011-10-08&   842.250&   842.253&   1.03&   0.67&          16&           3\\
 2011-10-09&   843.203&   843.207&   1.12&   0.56&          17&           3\\
 2011-10-10&   844.288&   844.292&   1.01&   0.83&          15&           3\\
 \hline
 \multicolumn{7}{c}{$K_s$-band}\\
\hline
 2011-09-08&   812.240&   812.244&   1.36&   0.72&          17&           3\\
 2011-09-27&   831.240&   831.243&   1.11&   0.42&          17&           3\\
 2011-10-03&   837.422&   837.426&   1.42&   0.62&          17&           3\\
 2011-10-04&   838.228&   838.232&   1.09&   0.40&          17&           3\\
 2011-10-05&   839.228&   839.232&   1.08&   0.61&          17&           3\\
 2011-10-09&   843.207&   843.211&   1.11&   0.56&          17&           3\\
 2011-10-10&   844.292&   844.296&   1.01&   0.83&          17&           3\\
\enddata
\tablecomments{MJD: Modiﬁed Julian Date (JD-2,455,000.5). AM: Median Airmass, IQ: Median image quality (in arcseconds) measured by the queued service observing at the CFHT. $N_f$ : Number of dithered frames. ET: Exposure time (in seconds) for each dithered frame. }
\end{deluxetable}

Optical images were obtained using the MegaCam detector, which has 36 CCDs with a small gap of $13\arcsec$ between CCDs in the region of interest. Each CCD is composed of $2048 \times 4612$ pixels, and a pixel scale of $0.187\arcsec$ pixel$^{-1}$ results in a full $\sim1\times1$ square degree field-of-view (FoV). The pre-processed images were downloaded from the \texttt{Elixir}{\footnote{\url{https://www.cfht.hawaii.edu/Instruments/Imaging/Megacam/dataprocessing.html}}} pipeline at CFHT. The pipeline performs the detrending of MegaCam images which includes the bad-pixel mask, the bias subtraction, the flat-field correction, and the elimination of the over-scan region \citep{magnier2004}. For each pre-processed image, we only analyzed 6 CCDs ($\#13, 14, 15, 22, 23,$ and 24) covering the region around the center of the cluster. 

We performed astrometric and photometric calibration of 6 CCDs at each epoch using \texttt{SCAMP} \citep{bertin2006} which compares an input source catalog generated with \texttt{SExtractor} \citep{bertin1996} to the reference catalog from the Sloan Digital Sky Survey \citep[SDSS,][]{alam2015}. With astrometric solutions from the \texttt{SCAMP} output, a mosaic image covering $\sim19'\times 19'$ region around the cluster center was created for each dithered frame using \texttt{SWARP} \citep{bertin2002}. Figure~\ref{fig:fov} displays one of the best-seeing (image quality $=0.51\arcsec$) dithered image mosaics in the $g$-band. On average, 7 dithered images were taken in each optical band within a typical observing sequence of $\sim 5$~minutes at each epoch. We extracted 112 $g$- and 93 $i$-band dithered images for our analysis which correspond to 16 and 14 epochs in the $g$ and $i$ bands, respectively. Table~\ref{tbl:log} summarizes the observations used in each epoch in our analysis.

NIR images were downloaded from the \texttt{$'$I$'$iwi}{\footnote{\url{https://www.cfht.hawaii.edu/Instruments/Imaging/WIRCam/IiwiVersion2Doc.html}}} (\texttt{IDL} Interpretor of the WIRCam Images) pre-processing pipeline at CFHT. These images were taken using WIRCam, which is a $2\times2$ array of four $2048\times2048$ HgCdTe HAWAII-RG2 detectors (pixel scale $\sim0.3\arcsec$ pixel$^{-1}$) with gaps of $45\arcsec$ between adjacent detectors, thus covering a FoV of $\sim21'\times 21'$. The \texttt{$'$I$'$iwi} pipeline detrends the data (dark subtraction and flat-fielding), subtracts the sky, and provides calibrated WIRCam data products. For WIRCam images, the center of M15 was placed at the center of detector 4 and the photometry was performed only on this detector (see Figure~\ref{fig:fov}). Multiple large WIRCam dithers cover a total region of $\sim12'\times 12'$ around the center of the cluster. Instead of co-adding all the dithered frames per epoch, we performed photometry on each dithered image separately. In each NIR band, 17 dithered frames per epoch were taken within a total observation time of $\sim7$ minutes. We obtained 178 images at $J$ and $119$ at $K_s$ for a total of 11 epochs at $J$ and 7 at $K_s$. Table~\ref{tbl:log} includes summaries for the NIR observations.

 We also created an astrometrically calibrated median-combined image from the dithers obtained in the best-seeing epoch as a reference image for optical and NIR bands separately. For this purpose, a weight map was created to mask bad pixels using \texttt{WeightWatcher} \citep{marmo2008}. The dithered images were calibrated astrometrically using an input source catalog generated with \texttt{SExtractor} together with the astrometric information from 2MASS in the \texttt{SCAMP}. The astrometric calibration has a root-mean-square (rms) error of only $\sim0.1\arcsec$ both internally between different images and externally with 2MASS. Finally, we used \texttt{SWARP} to produce a median-combined image by co-adding the dithered images at the instrument pixel scale. 

\subsection{Point-spread function photometry}

We obtained point-spread function (PSF) photometry on each image using \texttt{DAOPHOT/ALLSTAR} \citep{stetson1987} and \texttt{ALLFRAME} \citep{stetson1994} routines applied to each filter separately. For all point sources with brightness above $5\sigma$ of the detection threshold, a median full-width at half maximum (FWHM) was obtained using \texttt{SExtractor} on each image. As a first step, all sources above a $5\sigma$ detection threshold were found using \texttt{DAOPHOT} with FWHM as input, and aperture photometry was performed within a 3-pixel aperture. Next an empirical PSF was determined from up to 200 bright and isolated stars in each image excluding the sources within 700 pixels ($\sim2.2\arcmin$) for optical and 300 pixels ($\sim1.5\arcmin$) for NIR images from the crowded center of the cluster. A variable PSF was modeled as a Gaussian profile that varies quadratically with the position in the frame. For all sources with aperture photomery in the first step, the PSF photometry was performed using \texttt{ALLSTAR}. 

All the above steps were also performed on the reference co-added images, and a star-list was created for optical and NIR filters, separately. Frame-to-frame coordinate transformations were derived with respect to the reference image for all epoch images using \texttt{DAOMATCH/DAOMASTER}. The reference-star list and the derived transformations were used as input for the PSF photometry to \texttt{ALLFRAME}, which performs profile fitting of all sources in the reference-star list across all the frames, simultaneously. From the output \texttt{ALLFRAME} photometry, we selected 50 secondary standard stars which are bright and isolated and are outside the crowded center of the cluster (as defined for the optical and NIR above). These secondary standards were selected to have small photometric uncertainties ($<0.005$~mag), no epoch-to-epoch variability (rms $<0.01$~mag), and be present in all of the frames. Our secondary standards were used as input in the \texttt{TRIAL} program (provided by Peter Stetson) to correct the \texttt{ALLFRAME} photometry for the frame-to-frame changes in the zero-points due to epoch-dependent variations. Finally, \texttt{TRIAL} provides internal and external scatter, variability index, and the light curves of candidate variable stars.

\subsection{Photometric calibration}

\begin{figure}
\epsscale{1.2}
\plotone{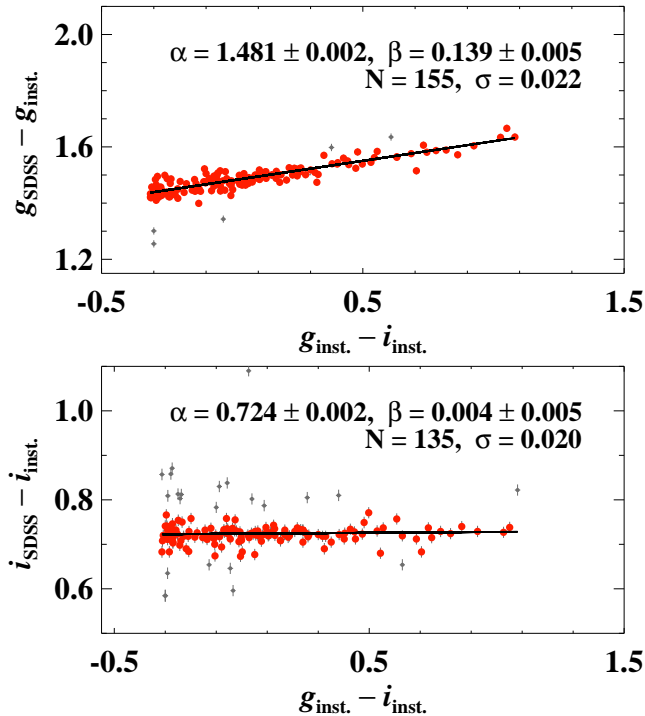}
\caption{Photometric transformations between the MegaCam instrument and the SDSS system. Each panel gives the coefficients of equation~(\ref{eq:sdss}). Small grey symbols represent $2.5\sigma$ outliers from the best-fitting linear regression.} 
\label{fig:sdss_cal}
\end{figure}

The astrometry for all sources in the optical and NIR catalogs were obtained from the calibrated median-combined reference images. The optical $gi$ photometry was cross-matched with the SDSS data release 12 catalogue \citep[][]{alam2015} within a tolerance of $1.0''$, which resulted in 1353 common stars. All the cross-matched sources with the SDSS are located on the outskirts of the cluster. We also applied different quality flags to select ``stellar objects'' with ``clean photometry'' and observations with ``good'' or ``acceptable'' quality. Furthermore, we restrict the sample to stars within the optimal magnitude range of $13<i_{\rm SDSS}<19$ and with uncertainties $<0.1$~mag. 

A final sample of 160 stars was used to derive photometric transformations from instrumental to SDSS magnitudes in the following form:

\begin{equation}
    g,i_{SDSS} - g,i_{inst.} = \alpha + \beta(g_{inst.}-i_{inst.}),
\label{eq:sdss}
\end{equation}

\noindent where $g,i_{\rm SDSS}$ are SDSS magnitudes in the $g$ and $i$ filters, and $\alpha$ and $\beta$ are the zero-points and color-coefficients, respectively. Note that an instrumental color-term was used in deriving photometric transformations because those exhibit significantly smaller uncertainties than the SDSS colors \citep[see Figure 1 of][]{an2008}.

Figure~\ref{fig:sdss_cal} shows the transformations and fitted coefficients for the calibration of instrumental magnitudes to SDSS $gi$ filters. A significant color term was obtained only for the $g$-band transformation, and the typical rms uncertainty of these transformations is only $\sim0.02$~mag. As an independent check, we also compared our calibrated photometry with the crowded-field photometry from \citet{an2008} for M15. A median offset of $\delta g =0.015$ mag and $\delta i= 0.004$ mag was obtained for all sources brighter than 18th mag which increases to $-0.024$ and 0.027 mag for all common sources within $1.0''$. 

 \citet{bhardwaj2020a} discussed in detail the photometric calibration of the WIRCam NIR data into the 2MASS system for the globular cluster M3. In brief, NIR photometry was cross-matched with the 2MASS catalog within a tolerance of $1.0\arcsec$ which resulted in 1684 common stars. This initial sample was restricted to stars with $11.5<J<16$~mag to avoid saturation and non-linearities at the bright end and larger uncertainties at the faint end. Finally, a clean sample of 322 stars was obtained for the calibration of the $J$ and $K_s$ data using sources with a photometric quality flag of ``A'' and those that are located outside 300 pixels from the cluster center. Absolute photometric calibration into the 2MASS system was obtained by correcting for a fixed zero-point offset ($J_{2MASS}-J_{inst.}=1.366~(0.005)$ mag and $K_s{_{2MASS}}-K{_{inst.}}=1.893~(0.004)$ mag, rms~$\sim0.05$~mag) that is independent of magnitude and color. We did not solve for a color-term because these 2MASS sources cover a very narrow range in color and can exhibit larger uncertainties (up to 0.15~mag). We estimate a systematic uncertainty of 0.03 mag for sources with $(J-K_s)<1.0$~mag while noting that all RR Lyrae stars have ($J-K_s$) colors smaller than 0.5 mag in M15.  

\section{Light curves of variable stars in M15}
\label{sec:var}

We adopted the list of variable star candidates including their coordinates, periods, types, and the pulsation modes from the catalog of \citet{clement2001}{\footnote{\url{http://www.astro.utoronto.ca/~cclement/}}}, which consists of 191 variables including 164 RR Lyrae and 3 Cepheid candidates. The catalog was last updated in 2014 and since then \citet{siegel2015} discovered an additional RRc variable. Several of these variable candidates have no accurate period determinations and have uncertain classifications. \citet{hoffman2021} improved the periods and classification of 79 candidate variables, although uncertainties remain for several candidates in the inner region of the cluster. 
The known variable candidate list was cross-matched with our optical and NIR data to extract light curves within an initial tolerance of $1\arcsec$. Since the adopted list of candidate variables includes positions of targets from different studies with different astrometric precisions, we increased the tolerance to $2\arcsec$ if no variable sources were found.  While all known variables except V104 and V105 are within the FoV of MegaCam, many of them fall within the gaps of $13\arcsec$ between the detectors (see Figure~\ref{fig:fov}). In the NIR data, five variables (V28, V43, V101, V104, V105) are outside the FoV of WIRCam. 

\begin{deluxetable}{rrrrr}
\tablecaption{Time-series photometry of RR Lyrae and Population II Cepheid variables in M15. \label{tbl:phot_lcs}}
\tabletypesize{\footnotesize}
\tablewidth{0pt}
\tablehead{\colhead{ID} & \colhead{Band} & \colhead{MJD} & \colhead{Mag.} & \colhead{$\sigma_{\textrm{mag}}$}}
\startdata
       V1&   $g$&    55778.477&    14.494&     0.003\\
       V1&   $g$&    56425.621&    14.634&     0.007\\
     ... &   ...&         ... &      ... &      ... \\
       V1&   $i$&    56415.625&    14.544&     0.004\\
       V1&   $i$&    56510.609&    14.711&     0.013\\
     ... &   ...&         ... &      ... &      ... \\
       V1&   $J$&    55843.203&    13.706&     0.021\\
       V1&   $J$&    55839.227&    13.947&     0.011\\
     ... &   ...&         ... &      ... &      ... \\
       V1& $K_s$&    55843.211&    13.472&     0.029\\
       V1& $K_s$&    55839.230&    13.579&     0.009\\
     ... &   ...&         ... &      ... &      ... \\
\enddata
\tablecomments{ID: `V'+ ID in the catalog of \citet{clement2001}; MJD $=$ JD $- 2,400,000.5$. The fourth and fifth columns represent magnitude and its associated uncertainty in the given band. This table is available in its entirety in machine-readable form. }
\end{deluxetable}

\subsection{Light curves and periods}

\begin{figure}
\epsscale{1.2}
\plotone{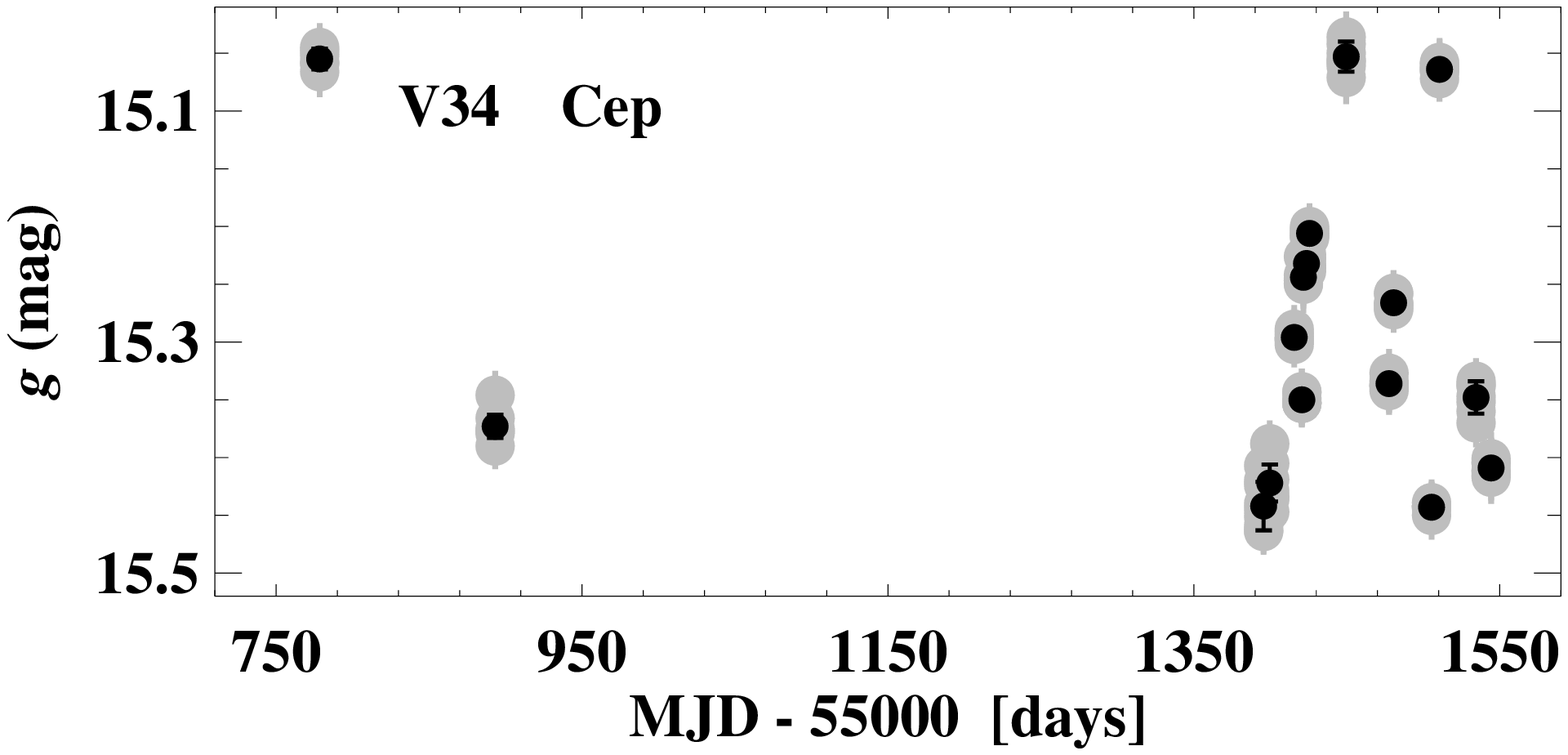}
\plotone{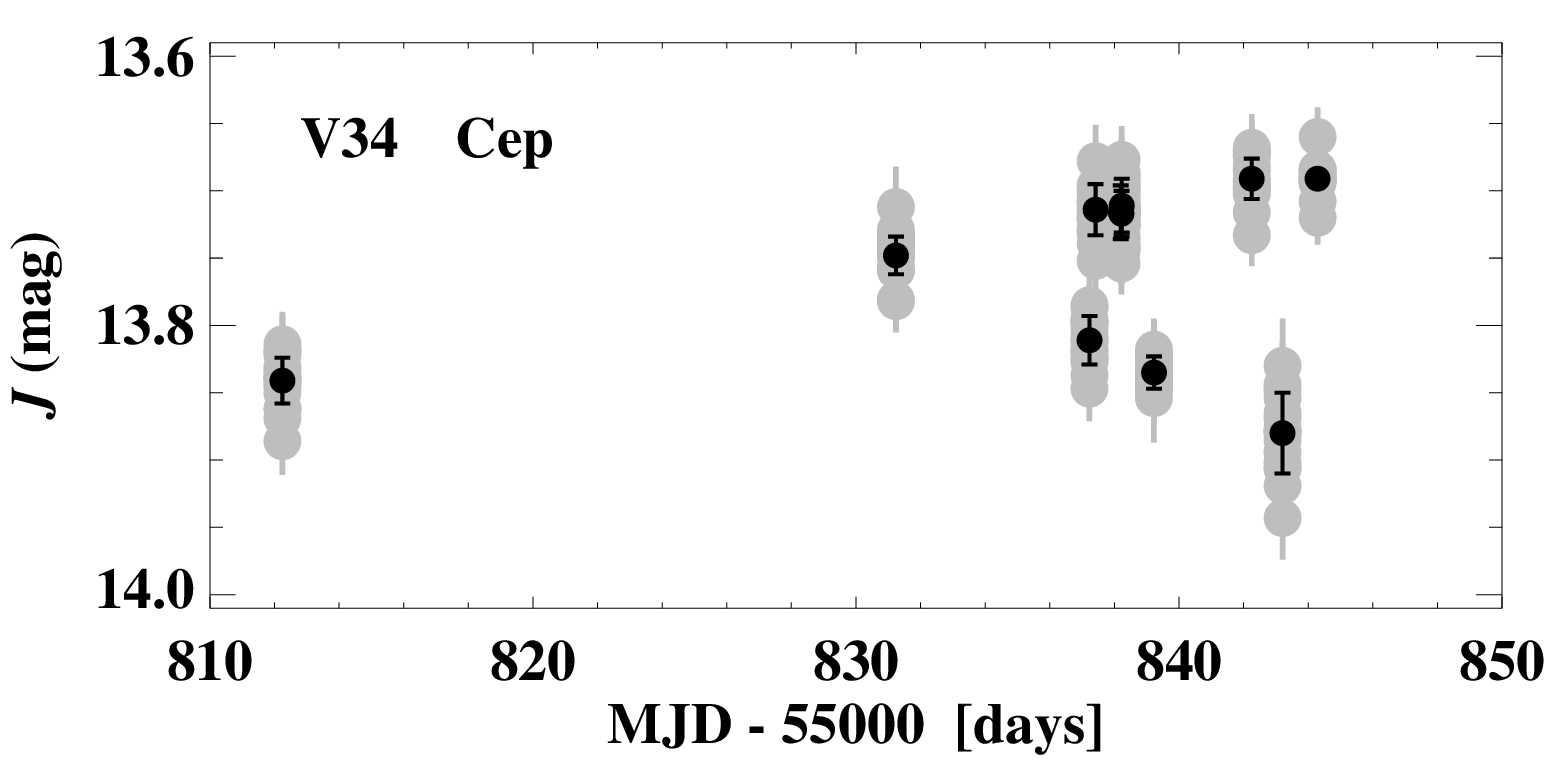}
\caption{Light curves of a candidate Cepheid variable star at $g$ (top) and $J$ (bottom). Grey symbols show all photometric measurements obtained from the dithered frames ($\sim$7 in $g$ and $\sim17$ in $J$) at a given epoch, and the black symbols represent the weighted mean magnitudes. } 
\label{fig:bin_lcs}
\end{figure}

\begin{figure}
\epsscale{1.2}
\plotone{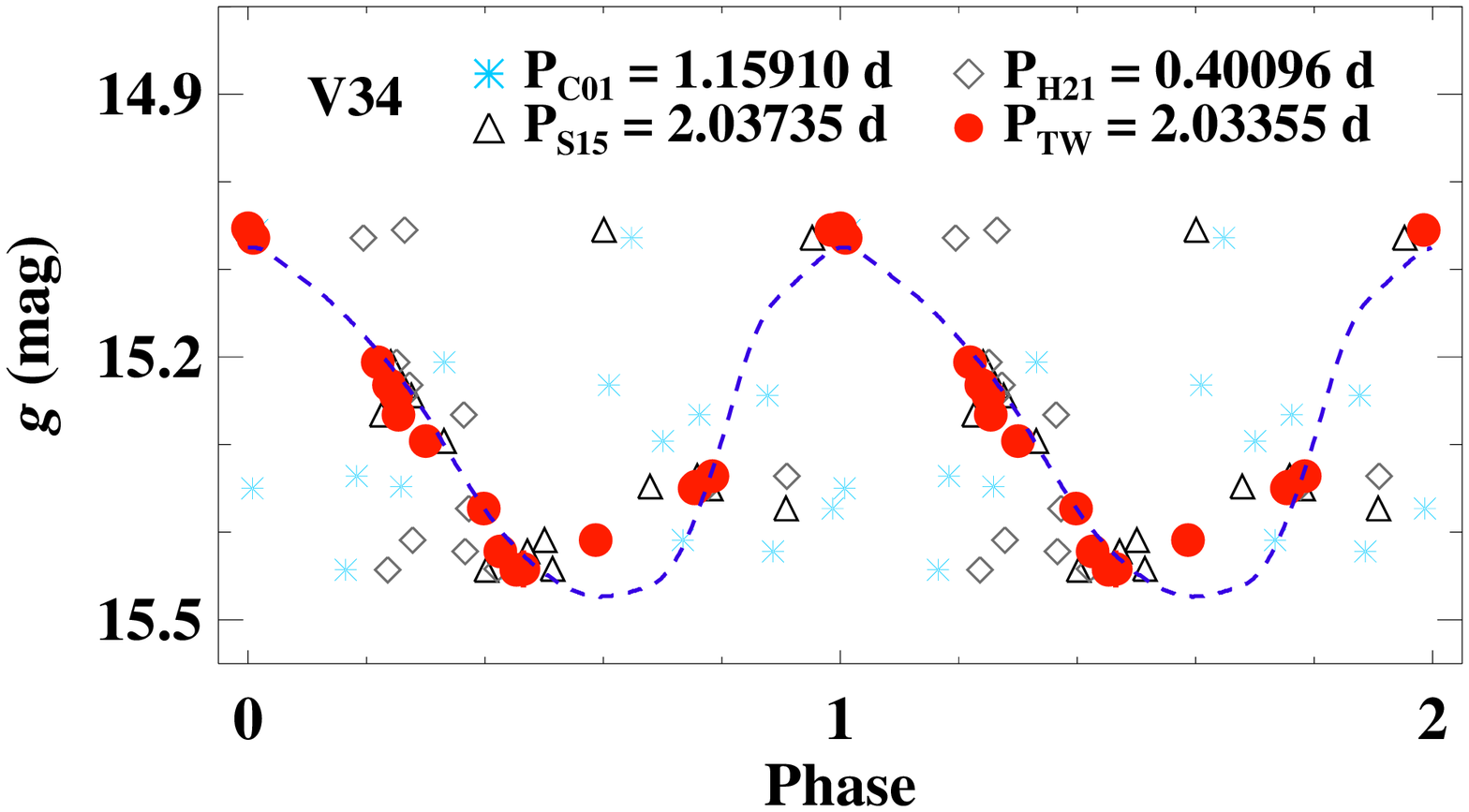}
\plotone{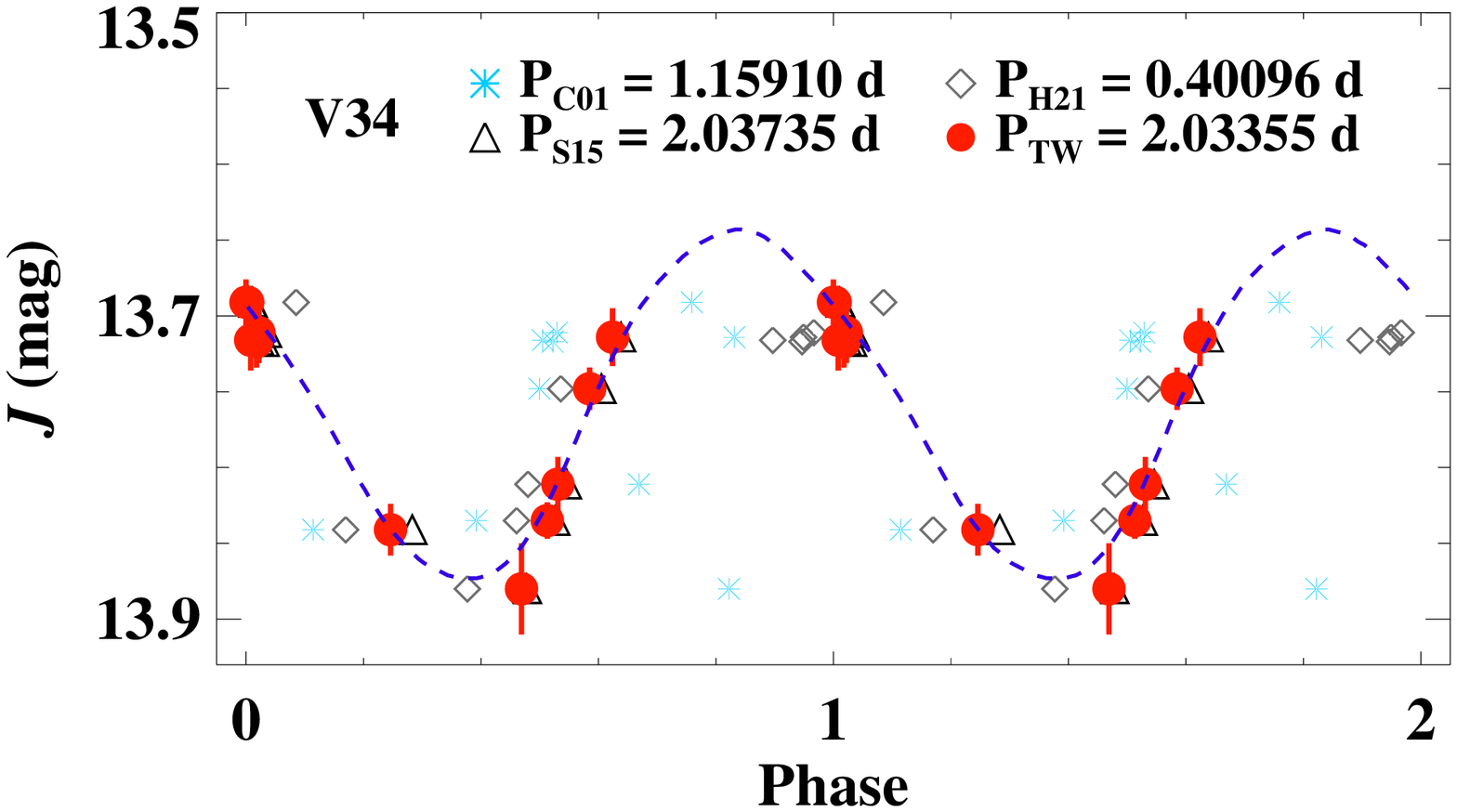}
\caption{Phased light curves of the same Cepheid candidate variable as shown in Figure~\ref{fig:bin_lcs} at $g$ (top) and $J$ (bottom) using different period determinations available in the literature. The periods are adopted from \citet[P$_{\rm C01}$,][]{clement2001}, \citet[P$_{\rm S15}$,][]{siegel2015}, \citet[P$_{\rm H21}$,][]{hoffman2021}, and our period (P$_{\rm TW}$) determined from the photometry in both optical bands.} 
\label{fig:per_lcs}
\end{figure}

Since the total integration times for all dithered frames in a given filter within an epoch are significantly smaller than the time scale of variability of our targets, all the photometric data points at a given epoch were binned to obtain optical and NIR light curves. We take a weighted mean of magnitudes obtained from all dithered frames at a given epoch and propagate the standard deviation of the robust mean to the photometric uncertainties. 

Figure~\ref{fig:bin_lcs} shows an example of binned light curve for a candidate Cepheid variable. The optical photometry has significantly smaller photometric uncertainties even for an individual dithered frame, and within a given epoch the weighted means are statistically accurate and precise. Since the individual WIRCam dithered frames have several bad pixels and the PSFs are relatively undersampled (see Table~\ref{tbl:log}), larger scatter is seen in the photometry at a given epoch. The light curves constructed from binned   photometry within a given epoch are used for further analysis in this work, but not for determining periods.

The long temporal baseline of our photometric data allows better constraints on the pulsation periods despite a cadence that at first glance does not appear to be suitable for short-term variability. We used the hybrid algorithm for period analysis from sparsely sampled multi-band data by \citet{saha2017} to determine pulsation periods between 0.01 and 100 days for all variable sources using optical data. The light curves were phased using our periods and those adopted from \citet{clement2001}, \citet{siegel2015}, and \citet{hoffman2021}, with the reference epoch corresponding to the maximum-brightness of our $g$-band observations. Figure~\ref{fig:per_lcs} displays the phased light curve of one of the candidate Cepheid variables with periods determined in other studies. Our derived period results in the minimum dispersion in the phased light curves in both optical and NIR filters.

We inspected all phased light curves visually and considered a star as variable if the ratio of scatter between different epochs and the photometric uncertainty exceeds the limit $\sigma_{\rm ext}/\sigma_{\rm phot} > 2.5$ and/or a clear periodicity is observed in the phased light curves. A fifth-order Fourier sine series \citep{bhardwaj2015} was fitted to estimate scatter in the light curves phased with periods from different studies. The best period is selected for all variables based on the minimum scatter in the phased light curve, and a final sample of 134 RR Lyrae and population II Cepheid variables (130 stars with optical light curves and 120 stars with NIR light curves) is selected for further analysis. We also recovered periodicity in 3 (V156, VZK68, and VNV1) out of 4 SX Phoenicis variables listed in the catalog of \citet{clement2001} in our optical data. The light curves of these SX Phoenicis and two candidate eclipsing binary stars (V157 and V158) exhibit large scatter due to relatively lower amplitudes and fainter magnitudes than RR Lyrae stars, and therefore, these variables are not analysed in this work. Table~\ref{tbl:phot_lcs} provides the time-series photometry for RR Lyrae and population II Cepheid variables in M15.
 
Table~\ref{tbl:m15_all} lists our adopted periods and those from \citet{clement2001}. In comparison to that work, five stars (V34, V72, V155, VZK3, and VNV11) have a period difference of greater than 0.1 days while 52 stars exhibit a difference $>10^{-4}$ days. For example, \citet{clement2001} does not list a period for V155, and \citet{hoffman2021} classified it as a Type II Cepheid with a period of 0.91189 days. However, their $V$-band light curve for V155 has a large scatter \citep[see Figure 6B,][]{hoffman2021}. Our good-quality optical and NIR light curves clearly show that V155 is an RRab star with a well-constrained period of 0.61251 days. Among the five stars showing a period difference of 0.1 days or greater, the largest difference was 1.115 days for VZK3.  Appendix \ref{sec:comments} discusses several RR Lyrae and population II Cepheid variables for which periods and classifications have improved significantly, and provides the reasons for excluding the remaining candidates from the list by \citet{clement2001} whose periodic and variable nature could not be confirmed in our photometry.

\subsection{Template fitting to the optical and NIR light curves}

\begin{figure}
\epsscale{1.2}
\plotone{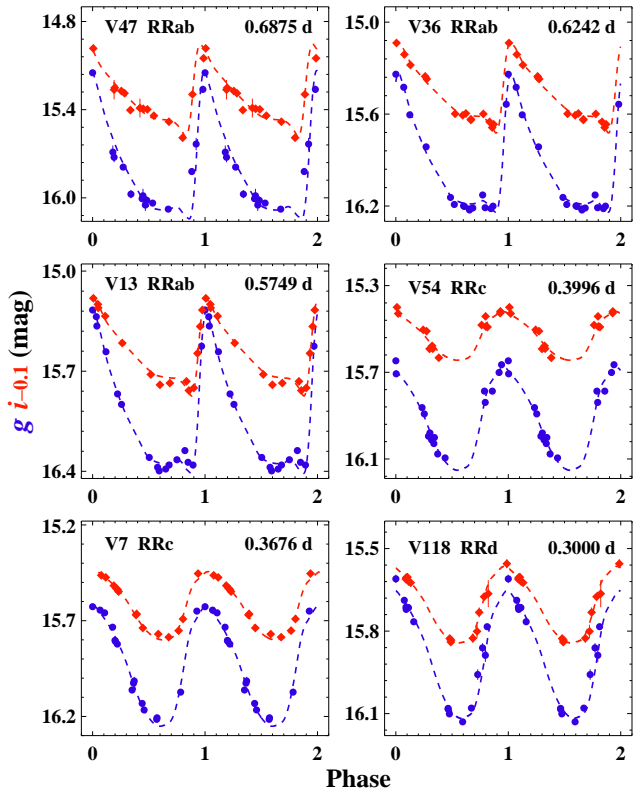}
\caption{Example phased light curves of RR Lyrae stars at $g$ and $i$ covering the entire range of periods in our sample. The $i$-band (red) light curves are offset for clarity by $-0.1$~mag only in this figure. The dashed lines represent the best-fitting templates \citep{sesar2010} to the data in each band. Star ID, variable subtype, and the pulsation period are included at the top of each panel.} 
\label{fig:lcs_gi}
\end{figure}

\begin{figure}
\epsscale{1.2}
\plotone{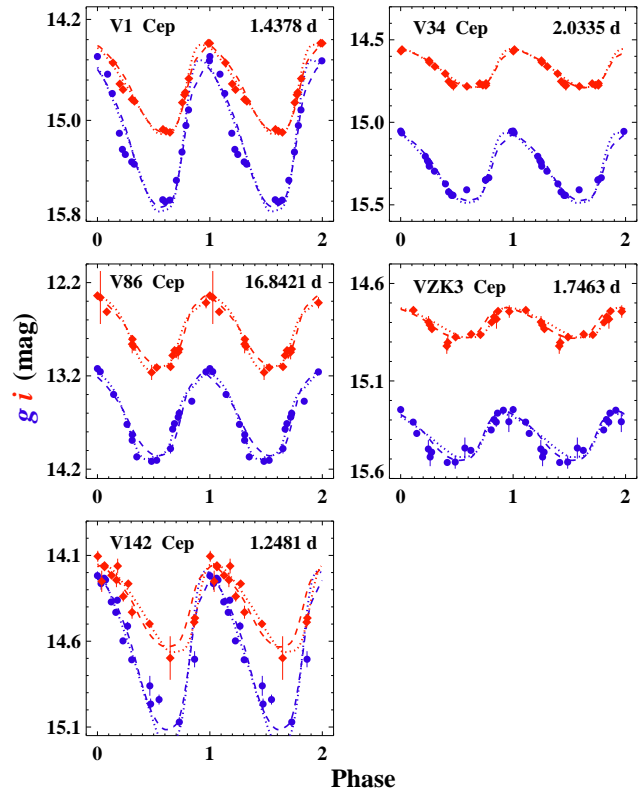}
\caption{Phased light curves of five Population II Cepheid candidates in $g$ and $i$. The dashed and dotted lines represent the best-fitting sinusoidal $gi$ RRc templates \citep{sesar2010} and $I$-band Type II Cepheid templates \citep{bhardwaj2017} to the data.} 
\label{fig:lcs_cep_gi}
\end{figure}

Optical light curves were fitted with templates in the $g$ and $i$ bands from \citet{sesar2010} based on RR Lyrae stars in the SDSS Stripe 82 region. The templates for Type II Cepheids are available only in the $I$ and $K_s$ bands \citep{bhardwaj2017}. Therefore, we fitted period-based $I$-band templates from \citet{bhardwaj2017} to both $gi$ light curves. Additionally, sinusoidal $gi$ RRc templates were also fitted in the case of Population II Cepheid variables. Since the reference epoch of maximum light is not well constrained, the templates were fitted to determine a mean magnitude, variable amplitude and a phase offset. The median magnitude and peak-to-peak amplitude from the phased light curves and a zero-phase offset were used as initial guesses to fit the template. In subsequent iterations, the phase offset was varied between $\pm$0.50 in steps of 0.02 in phase. Furthermore, we allow for up to $\pm20\%$ variation in the amplitude in steps of 0.02~mag. This tolerance ensures that the amplitudes are not underestimated in the cases with insufficient measurements around the maxima or minima and the amplitudes are not overestimated due to any outliers. The best-fitting templates based on the chi-squared minimization were adopted to determine optical mean magnitudes and amplitudes for RR Lyrae variables. In the case of Population II Cepheid variables, average mean magnitudes and amplitudes were derived from the two sets of templates applied to each light curve. Figs.~\ref{fig:lcs_gi} and ~\ref{fig:lcs_cep_gi} display the light curves and fitted templates to representative RR Lyrae stars with different periods and Cepheid variable candidates in M15, respectively. 

Similarly, NIR light curves were fitted with RR Lyrae templates from \citet{braga2019}. The $K_s$-band Type II Cepheid templates from \citet{bhardwaj2017} and near-sinusoidal $JK_s$ RRc templates \citep{braga2019} were used to fit the light curves of Cepheid variables. In the first pass, we fit templates for the mean magnitudes, variable amplitude and a phase offset as done for the optical light curves. However, the amplitudes are not well constrained due to the smaller number of epochs in the NIR data. Therefore, the median amplitude ratio ($\Delta J/\Delta g$ and $\Delta K_s/\Delta g$) obtained in the first pass was used to better constrain NIR amplitudes in the cases where light curves exhibit large phase gaps. As with the optical data, we also allowed for up to $\pm20\%$ variation in the amplitudes and also varied the phase offset to account for possible phase lag between optical and NIR light curves. 

\begin{figure}
\epsscale{1.2}
\plotone{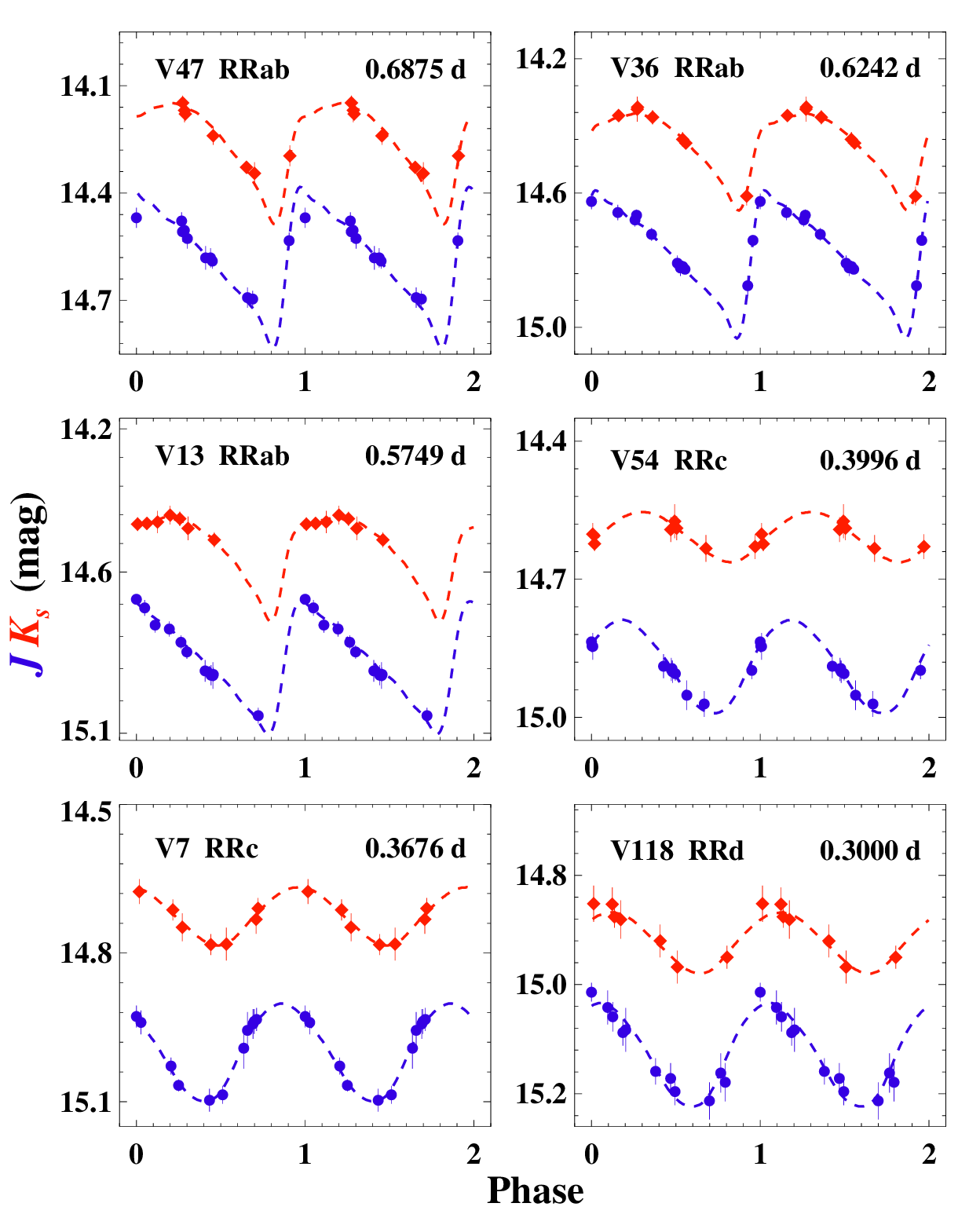}
\caption{Example phased light curves at $J$ and $K_s$ of the same RR Lyrae variables as in Figure~\ref{fig:lcs_gi}. The dashed lines represent the best-fitting templates \citep{braga2019} to the data in each band.} 
\label{fig:lcs_jk}
\end{figure}

\begin{figure}
\epsscale{1.2}
\plotone{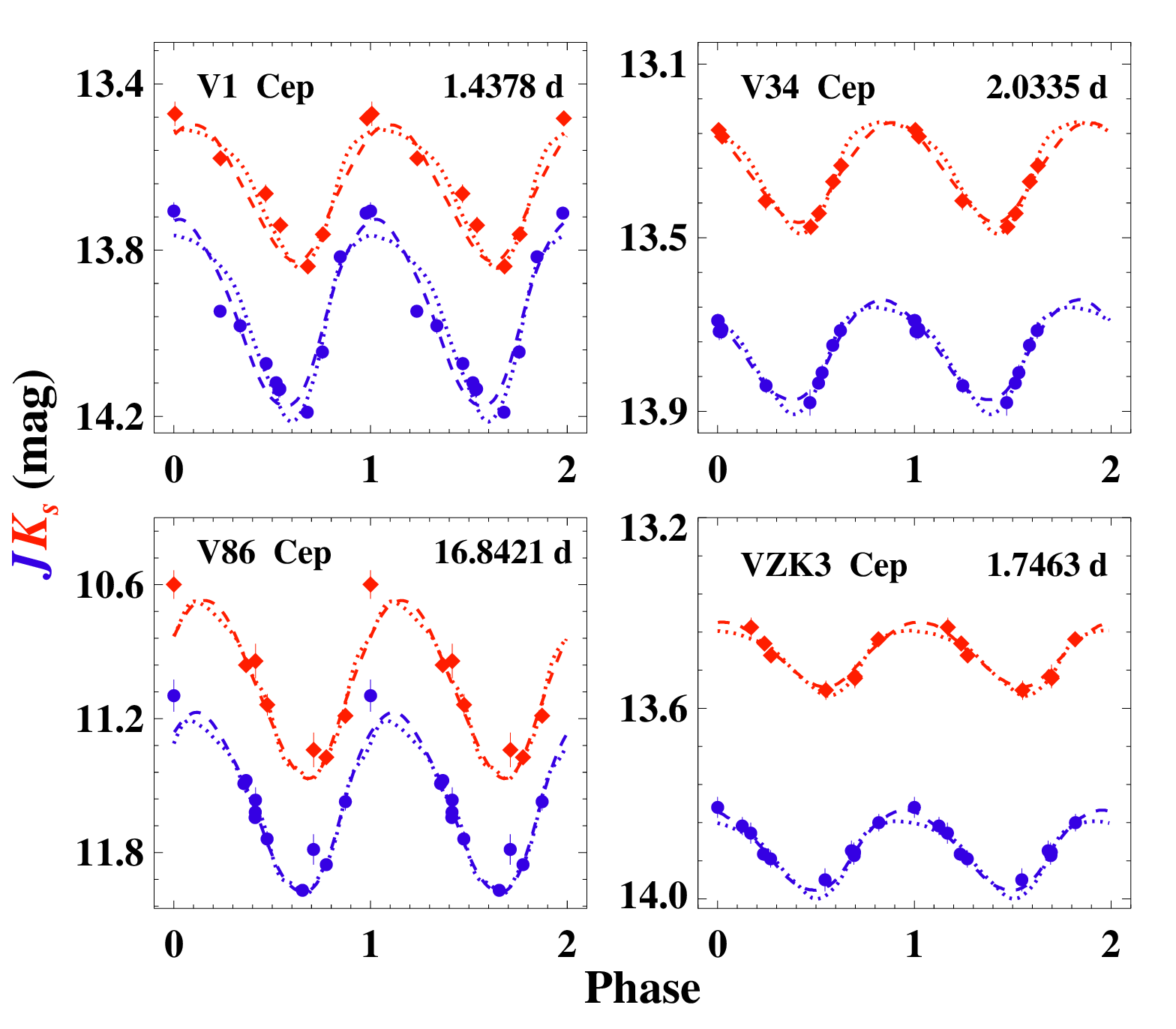}
\caption{Phased light curves of four (out of five) Population II Cepheid candidates from Figure~\ref{fig:lcs_cep_gi} at $J$ and $K_s$. The dashed and dotted lines represent the best-fitting sinusoidal $JK_s$ RRc templates \citep{braga2019} and $K_s$-band Type II Cepheid templates \citep{bhardwaj2017} to the data.} 
\label{fig:lcs_cep_jk}
\end{figure}

Furthermore, we fitted all three NIR RRab templates to the few variables where period-based templates did not fit all of the data points in the light curve to address evidence that the amplitude ratios change in different periods bins for clusters with different Oosterhoff types \citep{bhardwaj2020a}. Median photometric and the rms uncertainties of the templates were added to the errors on the mean magnitudes and amplitudes resulting from the best-fitting templates, respectively. Figs.~\ref{fig:lcs_jk} and ~\ref{fig:lcs_cep_jk} display NIR light curves and fitted templates to representative RR Lyrae stars with different periods and Cepheid variable candidates in M15, respectively. 

Table~\ref{tbl:m15_all} tabulates the optical and NIR pulsation properties including the intensity-averaged mean magnitudes and the peak-to-peak pulsation amplitudes of 129 RR Lyrae (51 RRab, 58 RRc, and 20 mixed-mode (RRd) variables) and 5 Population II Cepheid variable stars in M15. The limited temporal sampling in the CFHT data prevents us from identifying RR Lyrae stars that exhibit Blazhko variations. While previous studies of RR Lyrae stars in M15 have also not provided Blazhko classifications, these variations may well be present. For example, some Blazhko variations are apparent in the light curves at $B$ presented by \citet{corwin2008}.

\rotate
\begin{deluxetable*}{rccclccccccccccccccccc}
\tablecaption{Optical and NIR pulsation properties of RR Lyrae and Population II Cepheid variables in M15. \label{tbl:m15_all}}
\tabletypesize{\scriptsize}
\tablewidth{0pt}
\tablehead{\colhead{ID} & \colhead{RA} & \colhead{Dec} & \colhead{P$_{\rm C01}$} & \colhead{P$_{\rm TW}$} &  \colhead{Type}& \multicolumn{4}{c}{Mean magnitudes ($m_\lambda$)}  & \multicolumn{4}{c}{$\sigma_{m_\lambda}$} & \multicolumn{4}{c}{Amplitudes ($\Delta \lambda$) }  & \multicolumn{4}{c}{$\sigma_{\Delta \lambda}$} \\
 	&	&		& 	    &    &	   & $g$ & $i$  &   $J$  & $K_s$ & $g$ & $i$  &  $J$  & $K_s$  & $g$ & $i$  &   $J$  & $K_s$    & $g$ & $i$  &   $J$  & $K_s$	\\  
 	&	deg.&	deg.	& days	& days 	    &	   & \multicolumn{4}{c}{mag}  & \multicolumn{4}{c}{mag}  & \multicolumn{4}{c}{mag}   &  \multicolumn{4}{c}{mag}      }
\startdata
V1& 322.45917& 12.17414& 1.4377& 1.43781& BLH& 15.143& 14.762& 13.943& 13.649& 0.043& 0.044& 0.031& 0.032& 1.108& 0.684& 0.449& 0.329& 0.033& 0.029& 0.069& 0.067\\
V2& 322.44387& 12.16872& 0.6843& 0.68430$^1$& RRab& 15.981& 15.548& 14.724& 14.396& 0.020& 0.015& 0.025& 0.026& 0.699& 0.497& 0.341& 0.268& 0.027& 0.021& 0.040& 0.038\\
V3& 322.42229& 12.15394& 0.3887& 0.38873& RRc& 15.969& 15.731& 14.920& 14.643& 0.028& 0.020& 0.023& 0.022& 0.520& 0.287& 0.201& 0.125& 0.039& 0.027& 0.037& 0.032\\
V4& 322.46104& 12.12161& 0.3136& 0.31358& RRc& 15.887& 15.782& 15.142& 14.914& 0.026& 0.020& 0.024& 0.024& 0.667& 0.407& 0.202& 0.104& 0.036& 0.028& 0.040& 0.035\\
V5& 322.46471& 12.10811& 0.3842& 0.38421& RRc& 15.852& 15.631& 14.912& 14.642& 0.024& 0.014& 0.024& 0.024& 0.582& 0.324& 0.190& 0.129& 0.033& 0.019& 0.039& 0.036\\
V6& 322.49967& 12.18883& 0.6660& 0.66600& RRab& 15.910& 15.555& 14.738& 14.400& 0.039& 0.016& 0.025& 0.024& 1.027& 0.573& 0.377& 0.282& 0.053& 0.021& 0.037& 0.035\\
V7& 322.49575& 12.18800& 0.3676& 0.36756& RRc& 15.947& 15.727& 14.997& 14.723& 0.026& 0.015& 0.023& 0.023& 0.627& 0.356& 0.197& 0.118& 0.035& 0.021& 0.034& 0.034\\
V8& 322.49258& 12.20283& 0.6462& 0.64626& RRab& 15.958& 15.599& 14.770& 14.447& 0.033& 0.020& 0.025& 0.026& 1.041& 0.632& 0.383& 0.289& 0.045& 0.027& 0.036& 0.038\\
V9& 322.49683& 12.20608& 0.7153& 0.71528& RRab& 15.874& 15.480& 14.672& 14.333& 0.028& 0.018& 0.024& 0.024& 0.969& 0.511& 0.334& 0.271& 0.038& 0.025& 0.038& 0.036\\
V10& 322.52854& 12.16819& 0.3864& 0.38638& RRc& 15.952& 15.693& 14.931& 14.634& 0.028& 0.019& 0.022& 0.023& 0.540& 0.316& 0.169& 0.128& 0.039& 0.026& 0.037& 0.037\\
V11& 322.54167& 12.16181& 0.3433& 0.34322& RRc& 15.901& 15.749& 15.044& 14.789& 0.028& 0.021& 0.024& 0.023& 0.635& 0.363& 0.232& 0.160& 0.038& 0.029& 0.039& 0.042\\
V12& 322.53896& 12.15364& 0.5929& 0.59286& RRab& 15.945& 15.680& 14.826& 14.486& 0.037& 0.021& 0.029& 0.023& 0.913& 0.533& 0.382& 0.254& 0.051& 0.029& 0.042& 0.035\\
V13& 322.52896& 12.14869& 0.5749& 0.57491$^2$& RRab& 15.960& 15.687& 14.851& 14.539& 0.051& 0.034& 0.031& 0.023& 1.137& 0.681& 0.371& 0.292& 0.071& 0.047& 0.044& 0.035\\
V14& 322.51725& 12.09647& 0.3820& 0.38199& RRc& 15.953& 15.720& 14.943& 14.662& 0.019& 0.016& 0.024& 0.027& 0.588& 0.335& 0.166& 0.130& 0.025& 0.022& 0.038& 0.039\\
V15& 322.51654& 12.08319& 0.5835& 0.58364& RRab& 16.002& 15.682& 14.889& 14.592& 0.029& 0.020& 0.028& 0.027& 1.122& 0.754& 0.412& 0.308& 0.039& 0.026& 0.046& 0.045\\
V16& 322.52121& 12.20367& 0.3992& 0.39914& RRc& 15.943& 15.661& 14.908& 14.605& 0.018& 0.014& 0.024& 0.023& 0.524& 0.330& 0.155& 0.102& 0.024& 0.019& 0.038& 0.034\\
V17& 322.51646& 12.19822& 0.4294& 0.42900$^3$& RRd& 15.880& 15.576& 14.792& 14.481& 0.021& 0.011& 0.023& 0.023& 0.492& 0.299& 0.207& 0.127& 0.028& 0.015& 0.045& 0.033\\
V18& 322.51471& 12.19558& 0.3677& 0.36775& RRc& 15.909& 15.705& 14.964& 14.679& 0.031& 0.017& 0.023& 0.023& 0.624& 0.358& 0.201& 0.135& 0.042& 0.023& 0.038& 0.037\\
V19& 322.52421& 12.21228& 0.5723& 0.57231$^2$& RRab& 15.970& 15.668& 14.796& 14.536& 0.040& 0.024& 0.031& 0.026& 1.355& 0.847& 0.530& 0.272& 0.053& 0.032& 0.050& 0.041\\
V20& 322.51579& 12.16494& 0.6970& 0.69692& RRab& 15.936& 15.525& 14.671& 14.333& 0.033& 0.018& 0.025& 0.026& 0.918& 0.523& 0.378& 0.263& 0.044& 0.025& 0.037& 0.037\\
V21& 322.50250& 12.15153& 0.6476& 0.64882& RRab& 15.899& 15.532& 14.755& 14.438& 0.045& 0.026& 0.026& 0.026& 0.952& 0.641& 0.419& 0.326& 0.062& 0.036& 0.041& 0.038\\
V22& 322.39892& 12.15394& 0.7201& 0.72023$^2$& RRab& 15.891& 15.479& 14.707& 14.338& 0.019& 0.015& 0.031& 0.036& 0.768& 0.518& 0.338& 0.305& 0.025& 0.020& 0.046& 0.049\\
V23& 322.54675& 12.23889& 0.6327& 0.63270$^1$& RRab& 15.915& 15.543& 14.811& 14.440& 0.030& 0.027& 0.034& 0.035& 1.084& 0.585& 0.394& 0.327& 0.040& 0.038& 0.070& 0.056\\
V24& 322.46254& 12.16558& 0.3697& 0.36970& RRc& 15.920& 15.713& 14.994& 14.721& 0.025& 0.013& 0.023& 0.024& 0.590& 0.348& 0.210& 0.148& 0.034& 0.017& 0.037& 0.037\\
V25& 322.57875& 12.16503& 0.6653& 0.66532& RRab& 15.978& 15.575& 14.799& 14.470& 0.022& 0.013& 0.104& 0.035& 0.901& 0.519& 0.259& 0.215& 0.028& 0.017& 0.118& 0.047\\
V26& 322.49875& 12.25961& 0.4023& 0.40231$^3$& RRd& 15.998& 15.677& 14.928& 14.610& 0.017& 0.008& 0.026& 0.025& 0.414& 0.279& 0.184& 0.128& 0.024& 0.010& 0.045& 0.046\\
V28& 322.57967& 12.31644& 0.6706& 0.67060$^1$& RRab& 15.902& 15.629& ---& ---& 0.030& 0.019& ---& ---& 0.676& 0.524& ---& ---& 0.041& 0.026& ---& ---\\
V29& 322.53858& 12.22642& 0.5749& 0.57546& RRab& 16.091& 15.730& 14.947& 14.574& 0.033& 0.019& 0.037& 0.029& 0.786& 0.482& 0.292& 0.225& 0.045& 0.026& 0.055& 0.041\\
V30& 322.44596& 12.16611& 0.4060& 0.40600& RRd& 15.981& 15.704& 14.952& 14.646& 0.017& 0.011& 0.023& 0.023& 0.440& 0.199& 0.160& 0.127& 0.022& 0.015& 0.039& 0.037\\
V31& 322.46037& 12.23528& 0.4082& 0.40783$^3$& RRd& 15.962& 15.666& 14.950& 14.607& 0.017& 0.010& 0.023& 0.023& 0.433& 0.216& 0.158& 0.117& 0.022& 0.013& 0.037& 0.052\\
V32& 322.47829& 12.19728& 0.6044& 0.60551$^2$& RRab& 15.896& 15.564& 14.772& 14.471& 0.016& 0.011& 0.027& 0.024& 0.547& 0.316& 0.276& 0.214& 0.071& 0.065& 0.041& 0.037\\
V33& 322.48129& 12.15947& 0.5839& 0.58394$^3$& RRab& 15.979& 15.671& 14.891& 14.611& 0.045& 0.028& 0.037& 0.027& 1.191& 0.767& 0.473& 0.327& 0.062& 0.038& 0.049& 0.038\\
V34& 322.47721& 12.15208& 1.1591& 2.03355& BLH& 15.280& 14.678& 13.756& 13.340& 0.017& 0.019& 0.025& 0.026& 0.399& 0.225& 0.230& 0.230& 0.020& 0.013& 0.041& 0.037\\
V35& 322.48342& 12.12194& 0.3840& 0.38399& RRc& 15.961& 15.694& 14.946& 14.660& 0.023& 0.015& 0.024& 0.026& 0.543& 0.320& 0.200& 0.135& 0.031& 0.021& 0.037& 0.038\\
V36& 322.48508& 12.14489& 0.6242& 0.62412& RRab& 15.948& 15.574& 14.785& 14.457& 0.033& 0.022& 0.025& 0.024& 0.912& 0.594& 0.441& 0.289& 0.046& 0.031& 0.037& 0.036\\
V37& 322.48575& 12.14600& 0.2878& 0.28751& RRc& 15.900& 15.823& 15.185& 14.972& 0.020& 0.011& 0.023& 0.022& 0.590& 0.323& 0.187& 0.122& 0.026& 0.015& 0.039& 0.035\\
V38& 322.49521& 12.12692& 0.3753& 0.37527& RRc& 15.887& 15.680& 14.952& 14.681& 0.025& 0.017& 0.023& 0.023& 0.620& 0.365& 0.238& 0.146& 0.034& 0.023& 0.036& 0.034\\
V39& 322.49879& 12.13297& 0.3896& 0.38955$^2$& RRd& 16.000& 15.723& 14.943& 14.655& 0.019& 0.010& 0.023& 0.022& 0.520& 0.204& 0.180& 0.135& 0.026& 0.013& 0.052& 0.041\\
V40& 322.53029& 12.13531& 0.3777& 0.37733& RRc& 15.932& 15.700& 14.938& 14.655& 0.021& 0.012& 0.023& 0.025& 0.603& 0.339& 0.216& 0.147& 0.029& 0.017& 0.037& 0.039\\
\enddata
\tablecomments{Star ID, coordinates (epoch J2000), period (P$_{\rm C01}$), variable type and the pulsation mode are adopted from \citet{clement2001}. The period (P$_{\rm TW}$) is from this work unless specified with the following notes: $^1$ - \citet{clement2001}, $^2$ - \citet{hoffman2021}, $^3$ - \citet{siegel2015}. Type - RRab: fundamental-mode RR Lyrae, RRc: overtone-mode RR Lyrae, RRd - mixed-mode RR Lyrae, BLH: BL Herculis, WVI: W Virginis, ACF - fundamental-mode anomalous Cepheid. Last 16 columns represent intensity-averaged mean magnitudes and their errors, and peak-to-peak amplitudes and their uncertainties in the $giJK_s$ bands, respectively. Only the first 40 out of 134 variables are listed here, and the table is available in its entirety in machine-readable form.}
\end{deluxetable*}

\section{Optical and NIR pulsation properties of RR Lyrae and Cepheid variables}
\label{sec:puls}

\subsection{Color--magnitude diagrams}

The optical and NIR photometry for M15 was used to investigate the location of variable stars on the color--magnitude diagrams. To clean the possible contamination from field stars, our photometric catalogs were cross-matched with {\it Gaia} EDR3 \citep[][]{lindegren2021}. Out of 39,209 point-like sources in the FoV of MegaCam, we found only 27,736 with a $1.0\arcsec$ matching radius. Our photometry also covers fainter targets that are below the detection limit of {\it Gaia} ($g\sim22$~mag). In the case of the NIR data, we found 14,654 stars (out of 16,682 point sources) common with the {\it Gaia} data. The mean proper motions along the right ascension and declination axes are $\mu_\alpha = -0.655\pm0.010$ and $\mu_\delta = -3.828\pm0.009$ mas yr$^{-1}$, and these agree with the {\it Gaia} DR2 results \citep{helmi2018}.

\begin{figure}
\epsscale{1.2}
\plotone{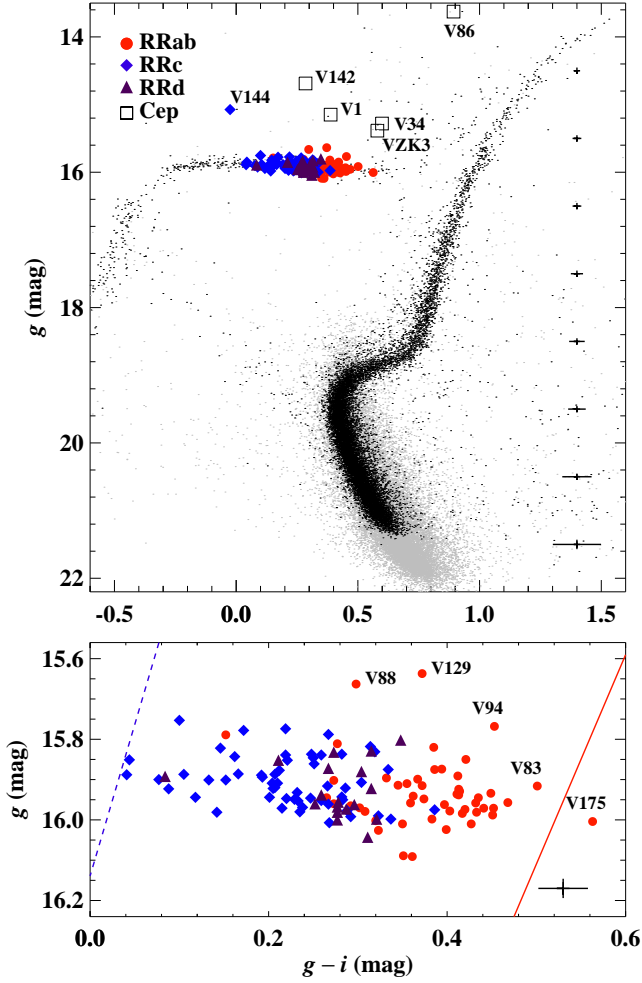}
\caption{\textit{Top}:  Variables in the optical color--magnitude diagram of M15. Some of the
variables above the horizontal branch are labeled.  Black symbols mark members of M15 based on their proper motions, and gray symbols mark the remaining point
sources in the FoV of MegaCam. The representative error bars are $\pm5\sigma$ in magnitude and color. \textit{Bottom}:  Close-up of the RR Lyrae stars on the horizontal branch. The dashed blue and solid red lines give the blue edge of the first overtone and the red edge of the fundamental mode, respectively. See
the text for details.} 
\label{fig:cmd_gi}
\end{figure}

The proper motions of all stars within the FoV of MegaCam and WIRCam in M15 were used to clean the optical and NIR color--magnitude diagrams, respectively. The proper motions along the right ascension and declination axes exhibit small spreads (0.57 and 0.50 mas yr$^{-1}$ half-width half-maximum) even though no astrometric quality cuts were applied. We removed all sources beyond $\pm5\sigma$ scatter around the mean proper motions. Among 134 variables analysed in this work, only seven candidates (V33, V47, V88, V94, V106, V168, VZK69) are $\pm5\sigma$ outliers in the proper motions along the right ascension and/or the declination axis. However, their proper motions exhibit very large uncertainties (renormalised unit weight error $>$ 5) indicating spurious astrometric solutions. Therefore, we consider all variables in our sample as members of the cluster. Figure~\ref{fig:cmd_gi} shows the optical color--magnitude diagram for the most probable members of M15 based on their proper motions. Variable sources are overplotted and the rich RR Lyrae population is evident on the horizontal branch. Five Cepheid candidates are marked which are significantly brighter than the horizontal-branch stars. One of the RRc variables (V144) is also bluer and brighter than the horizontal-branch stars. The optical $g$ and $i$-band light curves of V144 show clear periodicity and variability (e.g., $\Delta g\sim0.21$~mag), but the star is located in the central $1\arcmin$ region of the cluster and is probably blended with a nearby bright star. 

\begin{figure}
\epsscale{1.2}
\plotone{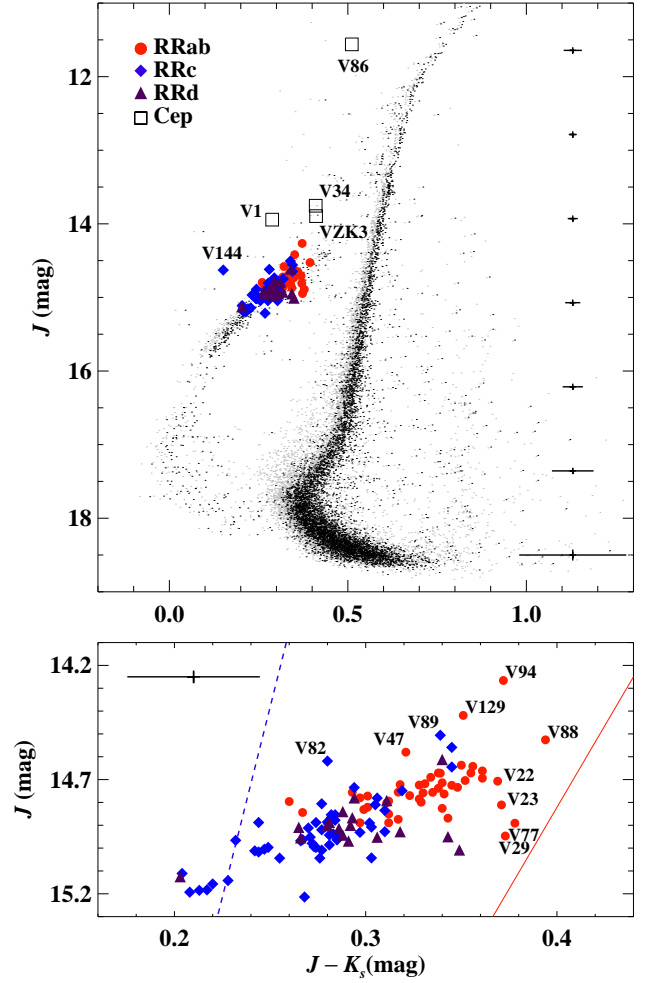}
\caption{\textit{Top}: Color--magnitude diagram in the NIR for M15, as observed with WIRCam.
\textit{Bottom}:  Close up of the RR Lyrae stars on the horizontal branch. Symbols and labels are as defined in Figure~\ref{fig:cmd_gi}.}
\label{fig:cmd_jk}
\end{figure}

The bottom panel of Figure~\ref{fig:cmd_gi} zooms in on the horizontal branch, where the RRab and RRc populations overlap significantly. The majority of mixed-mode variables fall within the so-called  ``either-or'' region \citep{vanalbada1973, stellingwerf1975, bono1995} in the instability strip where both pulsation modes are possible. \citet{marconi2015} provided analytical relations for the fundamental red edge and the first-overtone blue edge in $V-I,V$, which are transformed to $g-i,g$ in Figure~\ref{fig:cmd_gi}. We used the transformations for metal-poor Population II stars from \citet{jordi2006} to convert $V-I$ to $g-i$. For transforming $V$ to $g$, a median value of $(B-V)=0.4$ mag is adopted from \citet{corwin2006}, which can add a systematic uncertainty of up to 0.1 mag in transformed $g$ magnitudes. 

We also add an extinction to the transformed theoretical mean $g$ and $i$ magnitudes corresponding to the color-excess of $E(B-V)=0.1$ mag \citep{harris2010} towards M15. Total-to-selective absorption ratios in the $g$ and $i$ bands ($A_{g/i} = 3.79/2.09~E(B-V)$) were adopted from \citet{schlegel1998} for the extinction law of \citet{card1989} assuming $R_V=3.1$. Once the theoretical values are shifted for a true distance modulus of $15.15$ mag \citep{baumgardt2021} and corrected for an extinction corresponding to $E(B-V)=0.1$ mag, predicted boundaries are consistent with the observed distribution of RR Lyrae stars in M15. All of the RR Lyrae variables, except V175, are well within the theoretically predicted boundaries of the instability strip. 

The top panel of Figure~\ref{fig:cmd_jk} shows the NIR color--magnitude diagram for sources in M15. RR Lyrae and Cepheid variable candidates are also overplotted. Note that V144 is also an outlier in the NIR color--magnitude diagram, and its NIR light curves have small amplitudes (e.g. $\Delta J\sim0.08$~mag). These discrepancies, as in the optical, suggests that it is most likely brighter due to blending. We only found four (out of five in optical data) Cepheid candidates in NIR data because no good quality light curve was retrieved for V142. Three RRab stars (V88, V94, and V129) are also relatively brighter than the majority of RRab stars in both the optical and NIR color--magnitude diagrams and are located in the crowded center of the cluster. 

The bottom panel of Figure~\ref{fig:cmd_jk} displays the horizontal branch of M15 and the predicted boundaries of the instability strip. The metal-independent analytical relations for the instability strip boundaries in the $J,J-K_s$ color--magnitude diagram were taken from \citet{marconi2015}. The NIR extinction was also derived using $A_{J/K_s} = 0.95/0.38~E(B-V)$, assuming $R_V=3.1$ as with the optical data. Once distance and extinction corrections similar to the optical color--magnitude diagram are applied, most variables fall within the predicted boundaries of the instability strip. The variables that fall outside the predicted boundaries can be explained by the larger uncertainties in the NIR colors. The RRab and RRc variables overlap more in M15 than in the otherwise similar OoII type cluster M53 \citep{bhardwaj2021}. This suggests that most of the RR Lyrae may have evolved within the instability strip in M15 instead of their typical evolution from blue to the red edge in OoII clusters \citep{bingham1984,vandenberg2016}.

\subsection{Bailey diagrams and the amplitude ratios}

\begin{figure}
\epsscale{1.2}
\plotone{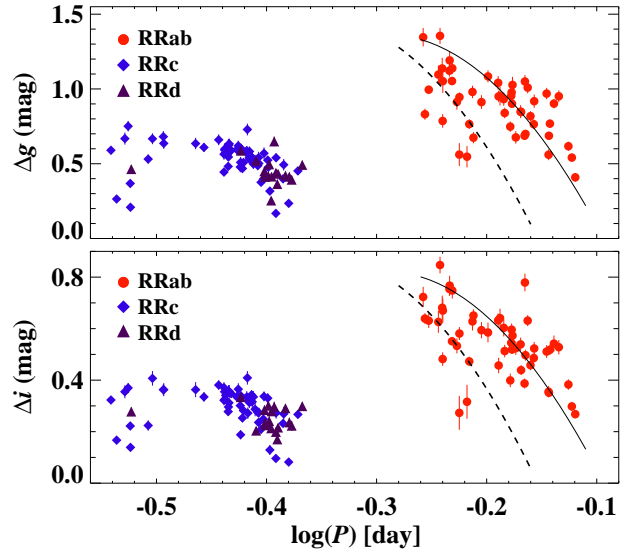}
\caption{Bailey diagrams for RR Lyrae stars in M15 in the $g$ (top) and $i$ (bottom) bands. In the case of the $g$ band, the dashed line displays the locus of OoI type RRab  based on the field RR Lyrae stars in the SDSS Stripe 82 region \citep{sesar2010}. The solid line is obtained by offsetting the OoI locus by $\log P=0.06$ \citep[as for M3 from][]{cacciari2005, bhardwaj2020a} and scaling it arbitrarily by $90\%$. For $i$, the loci from the top panel were scaled by $60\%$.}
\label{fig:bailey_gi}
\end{figure}

\begin{figure}
\epsscale{1.2}
\plotone{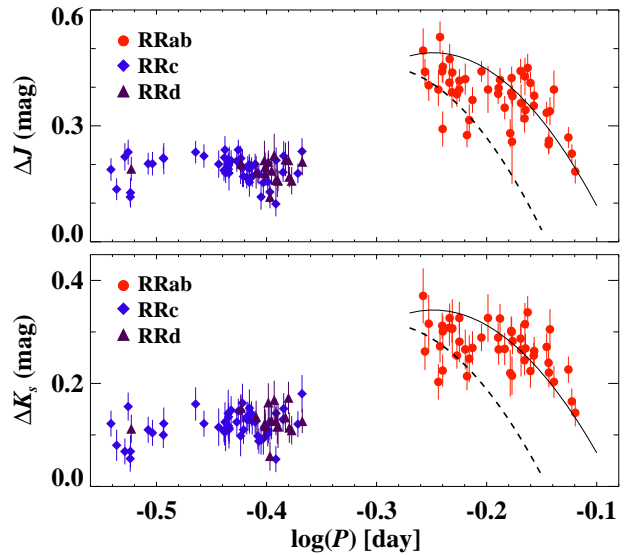}
\caption{Bailey diagrams for RR Lyrae stars in M15 at $J$ (top) and $K_s$ (bottom). In the top panel, dashed and solid lines display the locus of OoI and OoII type RRab in $J$ from \citet{bhardwaj2020a}. 
In the bottom panel, OoI and OoII loci of the $J$ in the top panels were scaled by $70\%$ as approximate loci in $K_s$.}
\label{fig:bailey_jk}
\end{figure}

\begin{figure}
\epsscale{1.2}
\plotone{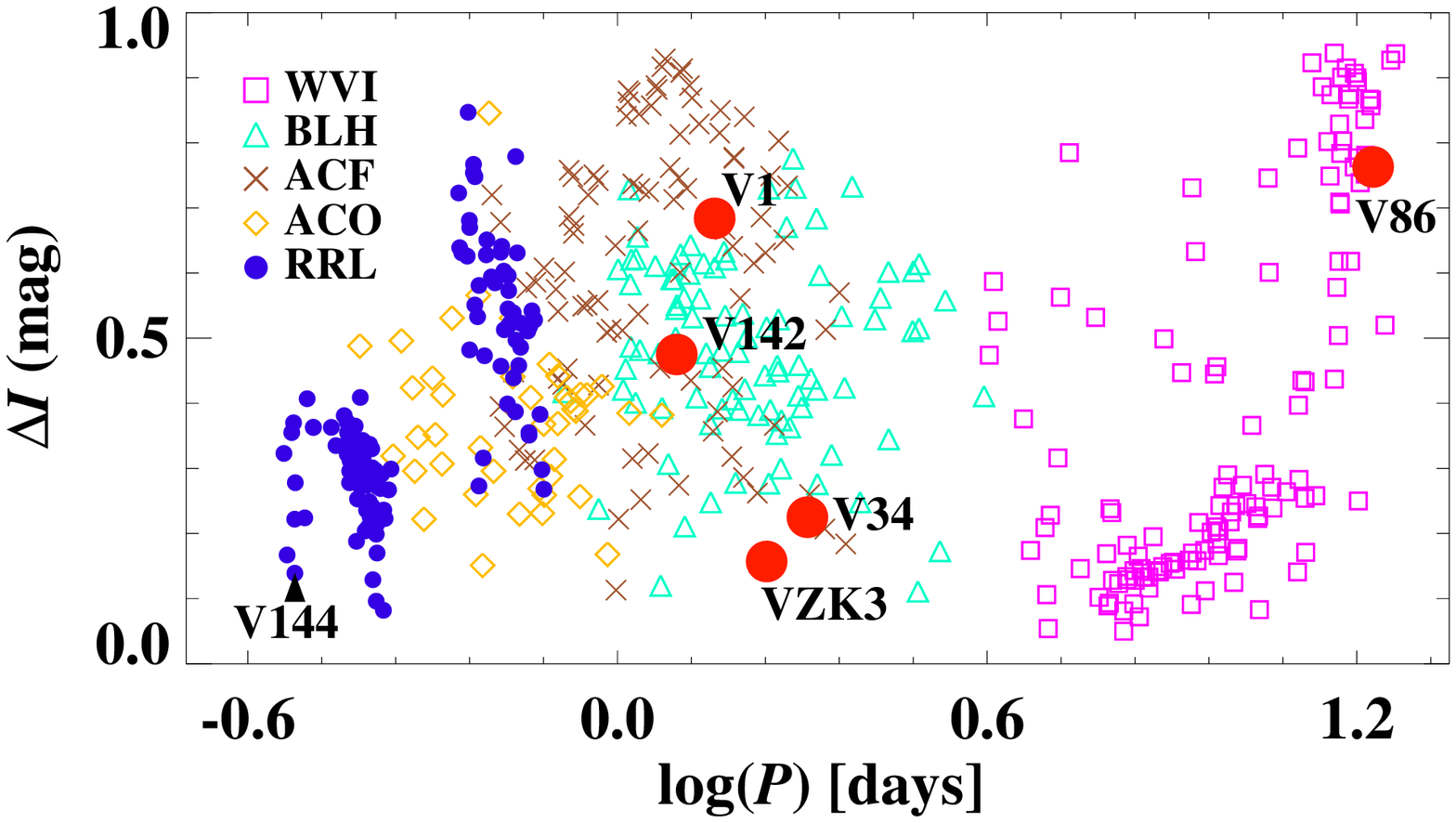}
\plotone{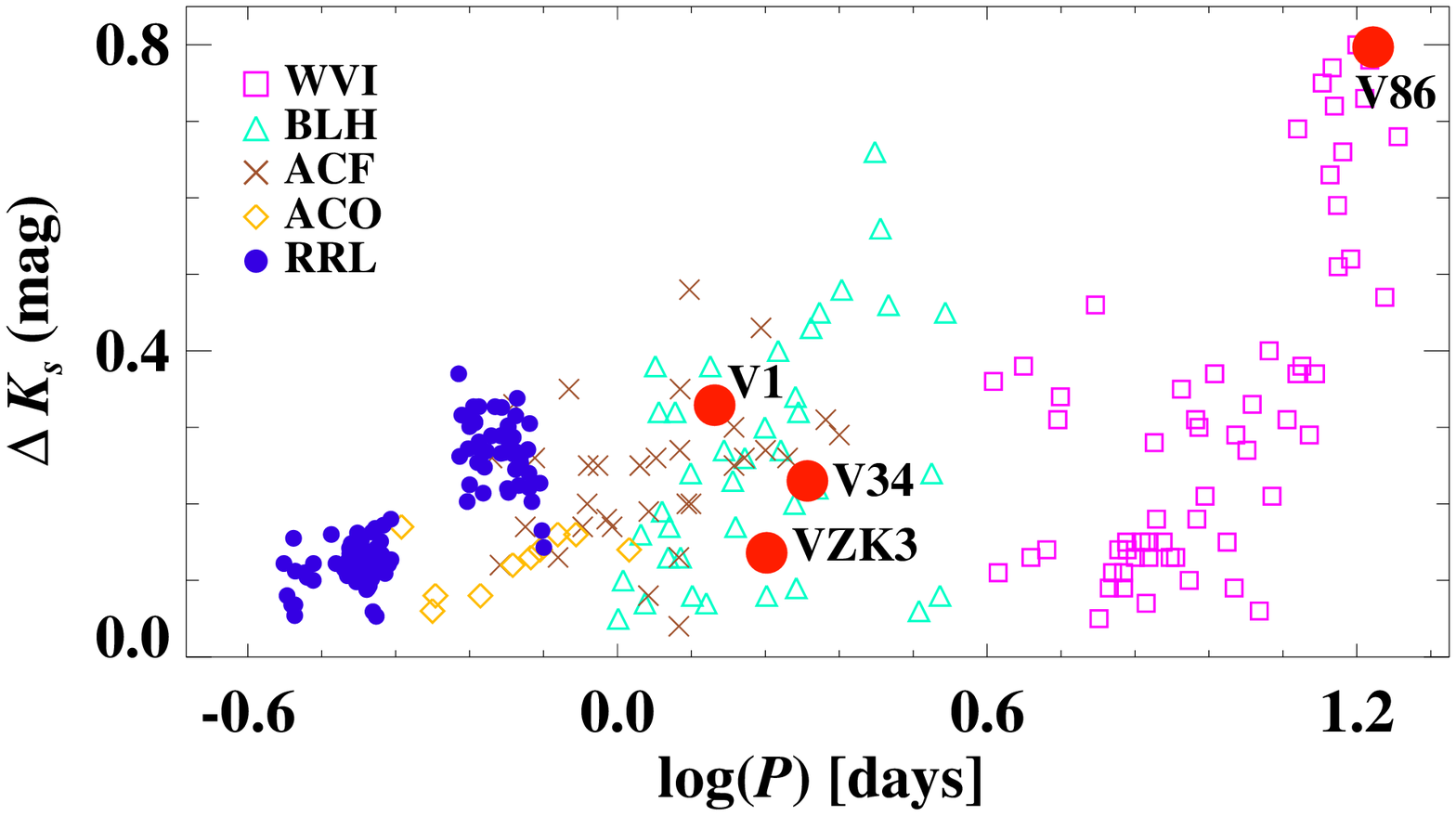}
\caption{Period--amplitude diagrams for Population II Cepheid and RR Lyrae stars in the $I$ (top) and $K_s$ (bottom) bands. The $I$-band amplitudes for BL Herculis (BLH), W Virginis (WVI), and fundamental and first-overtone mode anomalous Cepheids (ACF and ACO) are for the LMC variables \citep{soszynski2015a, soszynski2018}. RR Lyrae and Cepheid candidates (large red circles) are also shown. Optical amplitudes are in the OGLE $I$ band for the LMC variables but in SDSS $i$ for M15 variables. The $K_s$-band amplitudes are from \citet{bhardwaj2017} and \citet{ripepi2014a} for Type II Cepheid and anomalous Cepheids, respectively.}
\label{fig:pamp_ik}
\end{figure}

\begin{figure*}
\epsscale{1.2}
\plotone{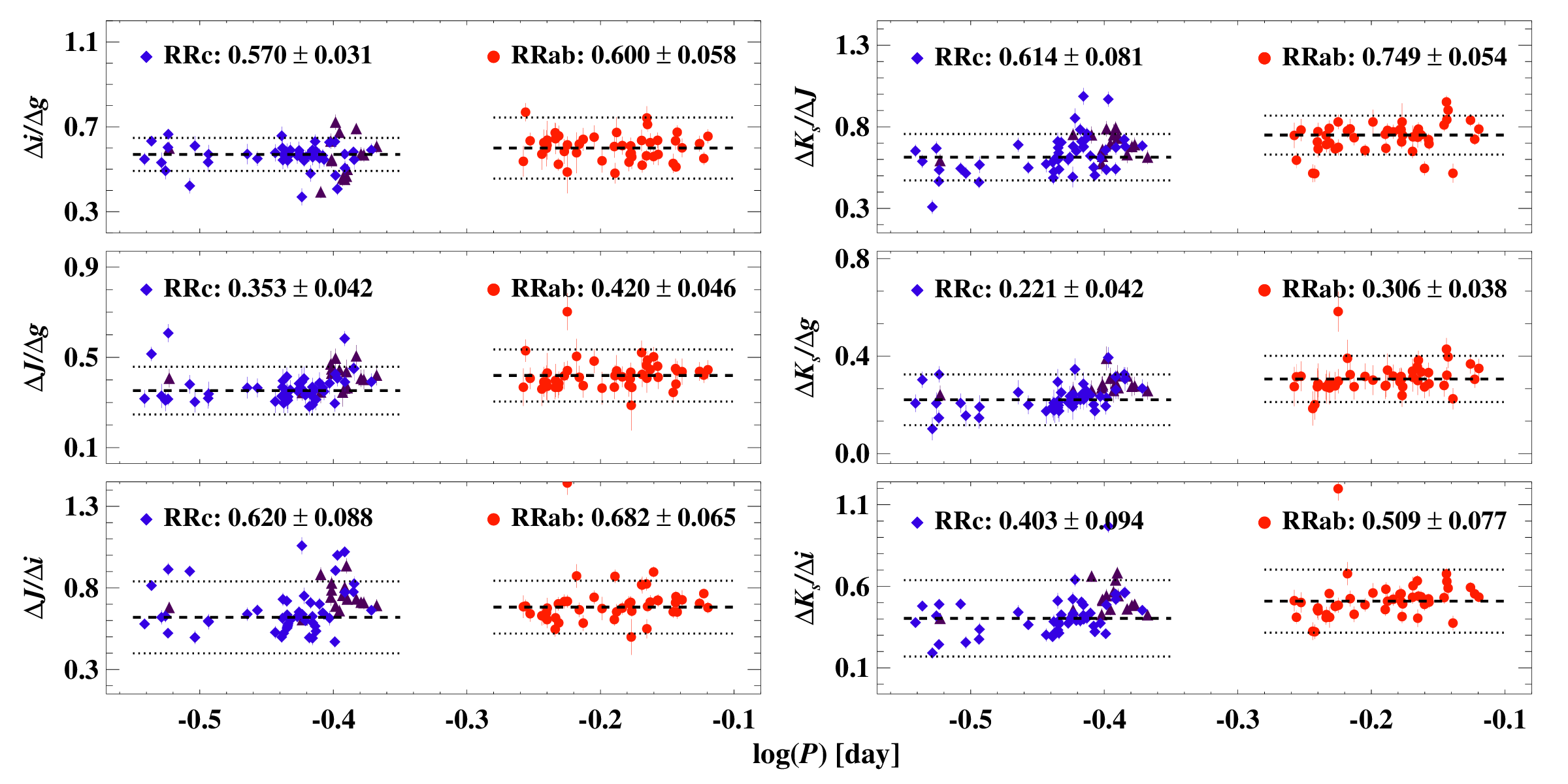}
\caption{Optical and NIR amplitude ratios for RR Lyrae stars in M15. The mean values and their standard deviations for RRc and RRab stars are given on the top of each panel. The dashed lines represent the mean values after iteratively removing $2.5\sigma$ outliers and the dotted lines represent $\pm2.5\sigma$ scatter around the mean values.}
\label{fig:amp_ratio}
\end{figure*}

The period--amplitude or Bailey diagrams for RR Lyrae stars \citep{bailey1902} provide insight into the Oosterhoff dichotomy in the GCs \citep{catelan2009, fabrizio2019}. The amplitudes of RRab decrease as the pulsation period increases while the amplitudes first typically rise and then fall for RRc stars as a function of period. Figure~\ref{fig:bailey_gi} shows the period--amplitude diagrams in the $g$ and $i$ bands for RR Lyrae stars. The optical amplitudes are well determined despite the relatively small number of epochs. For RRab stars, an expected decrease in amplitudes is evident with increasing period. Only a small fraction of RRab stars follow the OoI locus for field RR Lyrae variables in the SDSS Stripe 82 region derived by \citet{sesar2010}. The majority of RRab variables follow the OoII locus as their mean period (0.65 days) is also similar to RR Lyrae stars in other OoII clusters. 

Figure~\ref{fig:bailey_jk} displays the NIR period--amplitude diagram for RR Lyrae stars for the first time in M15. The loci of OoI and OoII type RRab stars in M3 \citep{bhardwaj2020a} are also shown for a comparison. Similar trends as in the optical Bailey diagrams are also seen despite the larger uncertainties in the NIR amplitudes. The Bailey diagrams in both the optical and NIR
hint at a separation in RRab population in two Oosterhoff groups around $\log P \sim-0.2$~days. However, as noted by \citet{corwin2008}, several RRab in M15 occupy an intermediate position in the Bailey diagram between OoI and OoII loci based on amplitudes determined from $B$-band light curves with excellent phase coverage. 

We also investigated the period--amplitude diagrams for Cepheid candidate variables as shown in Figure~\ref{fig:pamp_ik}. The amplitudes at $I$ are shown for the Type II Cepheids and anomalous Cepheids in the LMC, which are from the OGLE survey \citep{soszynski2015a, soszynski2018}. The amplitudes for M15 variables are in SDSS $i$ and may slightly differ from those in the $I$. The amplitudes in $K_s$-band for Type II Cepheids were adopted from \citet{bhardwaj2017} while those for anomalous Cepheids were taken from the VMC survey \citep{ripepi2014a}. The amplitudes for the longest-period Cepheid candidate (V86) are consistent with W Virginis variables in the LMC. However, the amplitudes for short-period Cepheid candidates (V1, V34, V142, VZK3) fall in the overlapping region for BL Herculis and fundamental-mode anomalous Cepheids. 

The amplitude ratios are useful to constrain the amplitudes for fitting templates if those are determined accurately in at least one filter. Figure~\ref{fig:amp_ratio} shows the amplitude ratios in any two bands based on our photometric light curves of RR Lyrae stars. No trend is seen in the amplitude ratios as a function of pulsation period, but robust mean values differ for RRab and RRc stars. \citet{braga2018} and \citet{bhardwaj2020a} found that the $\Delta JHK_s/\Delta V$ values vary for RRab stars at a break period of $\log P=-0.155$~days in $\omega$ Cen and $\log P=-0.222$ days in M3, respectively. However, contrary to $\omega$ Cen (also an OoII type GC), only a small fraction of RRab stars in M15 have periods longer than $\log P=-0.155$~days. The NIR amplitudes in M15 are less well constrained due to the smaller number of epochs (11/7 in $J/K_s$) with respect to those in M3, which had  $20~JHK_s$ epochs \citep{bhardwaj2020a}. Therefore, larger uncertainties in amplitudes resulting from the template fitting to sparsely sampled light curves may hide possible variations in amplitude ratios as a function of period. Nevertheless, the mean $\Delta K_s/\Delta J$ values are consistent with RR Lyrae stars in M3 and $\omega$ Cen and the typical trend of smaller amplitude ratios for RRc than RRab is also evident in all amplitude ratios.

\subsection{Period--luminosity relations} 
\label{sec:plr}

RR Lyrae stars are known to follow a visual magnitude--metallicity relation, and no significant dependence is seen on pulsation period in the $V$ band \citep{bono2003, muraveva2018}. However, a PLR is seen for wavelengths longer than $R$, which is due to the increased sensitivity of the bolometric correction to effective temperature at longer wavelengths \citep{catelan2004, marconi2015}. M15 gives us the opportunity to explore multiband PLRs for RR Lyrae stars in a metal-poor GC, which allows comparisons with PLRs in relatively metal-rich clusters. Therefore, we investigate the dependence of optical and NIR magnitudes on pulsation periods for RR Lyrae stars in M15. Optical and NIR mean magnitudes from our template-fitted light curves were used to derive PLRs of the following form:

\begin{equation}
{m_\lambda} = a_\lambda + b_\lambda\log(P),
\label{eq:plr}
\end{equation}

\noindent where $a_\lambda$ and $b_\lambda$ are the slope and zero-point of the best-fitting PLR for a given wavelength. The PLRs were derived separately for the sample of RRab and RRc stars and for a global sample of all RR Lyrae variables. We also fitted a single linear PLR to the entire sample after converting the periods of RRc variables from the first-overtone to the period corresponding to the fundamental mode using the equation: $\log(P_{\textrm{RRab}})=\log(P_{\textrm{RRc}})+0.127$ \citep{petersen1991, coppola2015}. Mixed-mode variables were treated as though they were RRc variables because of their dominant first-overtone mode pulsations. 

\begin{deluxetable}{rrrrrr}
\tablecaption{Optical and NIR PLRs of RR Lyrae in the M15 cluster. \label{tbl:plrs}}
\tabletypesize{\footnotesize}
\tablewidth{0pt}
\tablehead{\colhead{Band} & \colhead{Type} & \colhead{$b_\lambda$} & \colhead{$a_\lambda$} & \colhead{$\sigma$}& \colhead{$N$}\\}
\startdata
      $g$ &  RRab &    15.552$\pm$0.030      &     -0.111$\pm$0.160      &      0.040 &   41\\
     $g$ &   RRc &    15.520$\pm$0.061      &     -0.019$\pm$0.138      &      0.053 &   52\\
     $g$ &   All &    15.597$\pm$0.018      &      0.185$\pm$0.066      &      0.050 &  109\\
     $i$ &  RRab &    15.105$\pm$0.035      &     -1.292$\pm$0.184      &      0.053 &   46\\
     $i$ &   RRc &    14.906$\pm$0.050      &     -1.329$\pm$0.112      &      0.044 &   50\\
     $i$ &   All &    15.114$\pm$0.017      &     -1.222$\pm$0.060      &      0.047 &  111\\
\hline 
     $J$ &  RRab &    14.368$\pm$0.025      &     -1.564$\pm$0.126      &      0.036 &   42\\
     $J$ &   RRc &    13.935$\pm$0.051      &     -2.157$\pm$0.115      &      0.036 &   42\\
     $J$ &   All &    14.308$\pm$0.016      &     -1.855$\pm$0.057      &      0.040 &  102\\
   $K_s$ &  RRab &    14.005$\pm$0.023      &     -2.061$\pm$0.116      &      0.034 &   42\\
   $K_s$ &   RRc &    13.523$\pm$0.054      &     -2.615$\pm$0.121      &      0.038 &   42\\
   $K_s$ &   All &    13.948$\pm$0.015      &     -2.333$\pm$0.054      &      0.037 &  101\\
\enddata
\tablecomments{The zero-point ($b$), slope ($a$), dispersion ($\sigma$) and 
the number of stars ($N$) in the final PLR fits are tabulated.}
\end{deluxetable}

\begin{figure*}
\epsscale{1.15}
\plotone{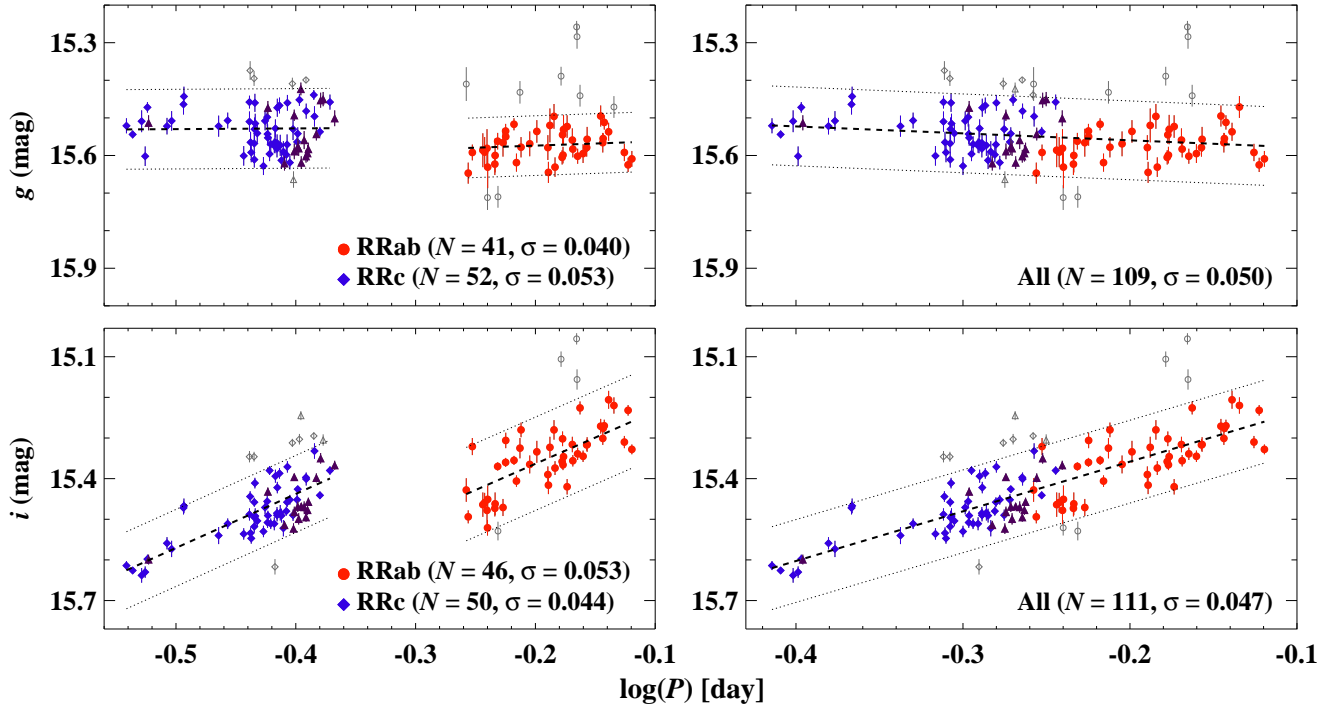}
\caption{Optical period-luminosity relations for RRab and RRc stars (left) and all RR Lyrae stars (right) in $g$ (top) and $i$ (bottom). In the right panels, the periods for the RRc/RRd variables have been shifted to their corresponding fundamental-mode periods, as explained in the text. The dashed lines represent best-fitting linear regressions over the period range under consideration while the dotted lines display $\pm 2.5\sigma$ offsets from the best-fitting PLRs.} 
\label{fig:gi_plr}
\end{figure*}

\begin{figure*}
\epsscale{1.15}
\plotone{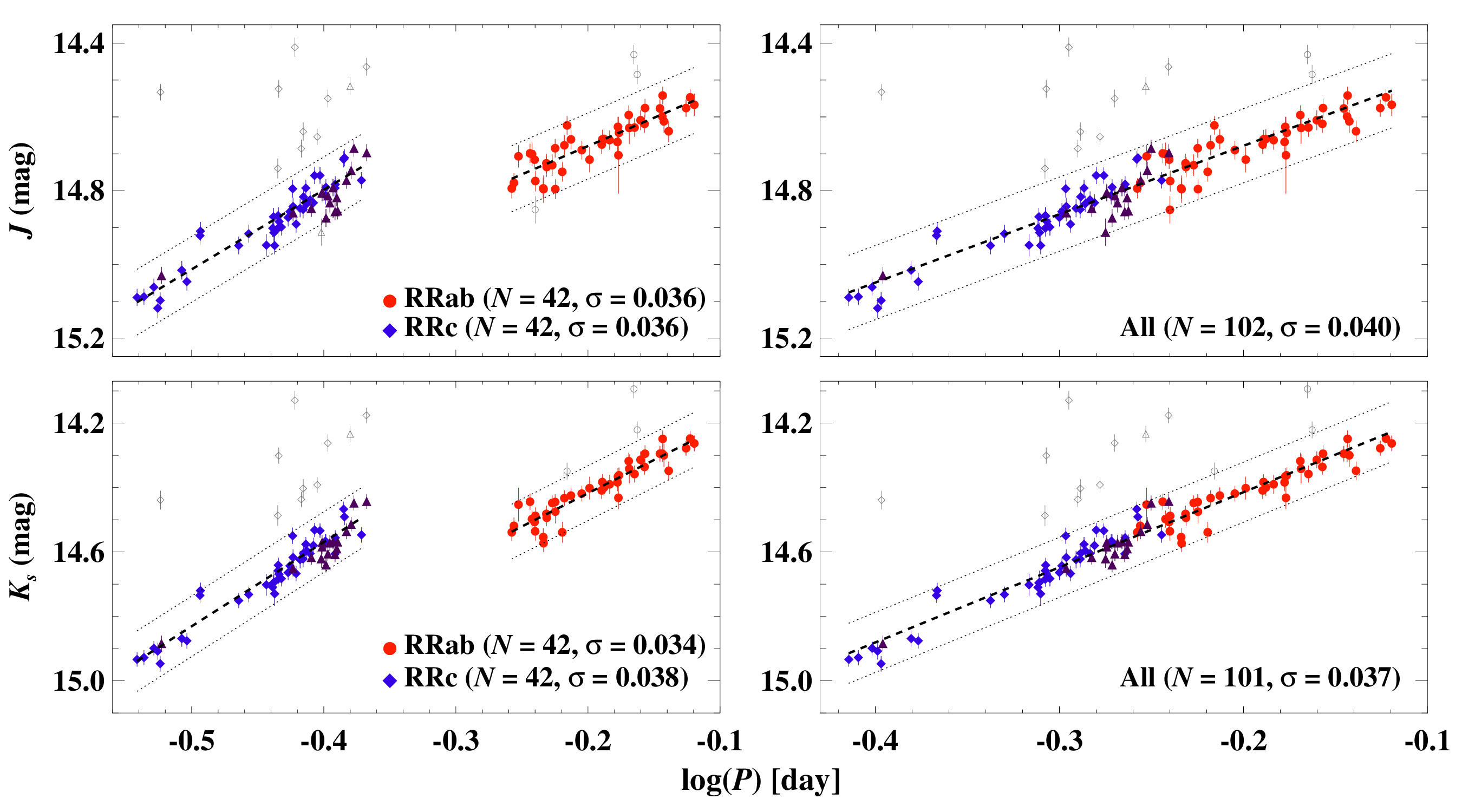}
\caption{NIR PLRs for M15 RRab and RRc (left) and all RR Lyrae (right) in $J$ (top) and $K_s$ (bottom). The periods for the RRc/RRd variables in the
right-hand panels have been shifted from the overtone to the fundamental mode. See Figure~\ref{fig:gi_plr} for an explanation of the plotted lines.} 
\label{fig:jk_plr}
\end{figure*}

Optical and NIR mean magnitudes were corrected for extinction before deriving the PLRs. The extinction corrections in each filter were applied based on the reddening value of $E(B-V) = 0.10$~mag \citep{harris2010} and the adopted extinction law of \citet{card1989} as discussed previously. Assuming $R_V=3.1$, the extinctions for all variables in M15 were $A_g$ = 0.379, $A_i$ = 0.209, $A_J$ = 0.095, and $A_{Ks}$ = 0.038 mag. 

Figure~\ref{fig:gi_plr} shows the optical PLRs for RRab, RRc, and all RR Lyrae stars in M15.  We iteratively removed the single largest outlier from an initial fit until all residuals were within $2.5\sigma$ scatter around the best-fitting relation. Table~\ref{tbl:plrs} presents the results of the best-fitting linear regression. Table~\ref{tbl:plrs} clearly shows that the slope of the $g$-band PLR is consistent with zero within $3\sigma$ of their quoted uncertainties, implying a small or negligible dependence on pulsation periods. This result agrees with the empirical relation at $g$ for RR Lyrae stars in M5 \citep{vivas2017} and theoretical predictions that show a tight RR Lyrae PLR only in the SDSS $i$ and $z$ bands \citep{marconi2006, caceres2008}. Indeed, the slope of the $i$-band PLR for the sample of all RR Lyrae stars is significantly steeper than at $g$ and it is statistically consistent with empirical relations in M5 \citep[For RRab: $-1.59$, for RRc: $-1.61$,][]{vivas2017} and the theoretical $i$-band PLZ relation \citep[RR Lyrae: $-1.04$,][]{caceres2008}. However, the dispersion in the PLRs at $g$ and $i$ is similar for different samples of RR Lyrae stars presumably due to the small number of epochs which limits the accuracy of mean magnitudes if the peak-to-peak amplitudes are not well-constrained.

Figure~\ref{fig:jk_plr} displays the NIR PLRs for RR Lyrae stars in M15. As shown first in the pioneering work of \citet{longmore1986}, the PLRs of RR Lyrae stars are tighter in the NIR. Table~\ref{tbl:plrs} shows that the RRc stars exhibit the steepest slope while the RRab variables show a shallower PLR, a trend that is commonly seen in other GCs (e.g., M3, M53, $\omega$ Cen) as well regardless of their Oosterhoff types. However, the global sample of RR Lyrae stars has a steeper slope than for just the RRab stars at both $J$ and $K_s$. \citet{bhardwaj2021} reported a similar result at $K_s$ in another OoII-type cluster, M53. The steeper slope becomes more evident if a stricter $2\sigma$ clipping threshold is adopted, which reduces the scatter to only $\sim0.02$~mag while retaining more than $70\%$ of the stars in all RR Lyrae samples. It is possible that the slopes for the global sample of RR Lyrae stars become steeper than those for the sample of only RRab stars in more metal-poor GCs ({[Fe/H]}$<-2.0$~dex). Adopting a different sigma threshold does not lead to statistically significant changes in either the slope or the zero-points of the PLRs. Furthermore, we also excluded the tail of the RRc stars ($\log P<-0.48$~days) and found that the coefficients of the PLRs are consistent with those listed in Table~\ref{tbl:plrs}. 

\begin{figure}
\epsscale{1.2}
\plotone{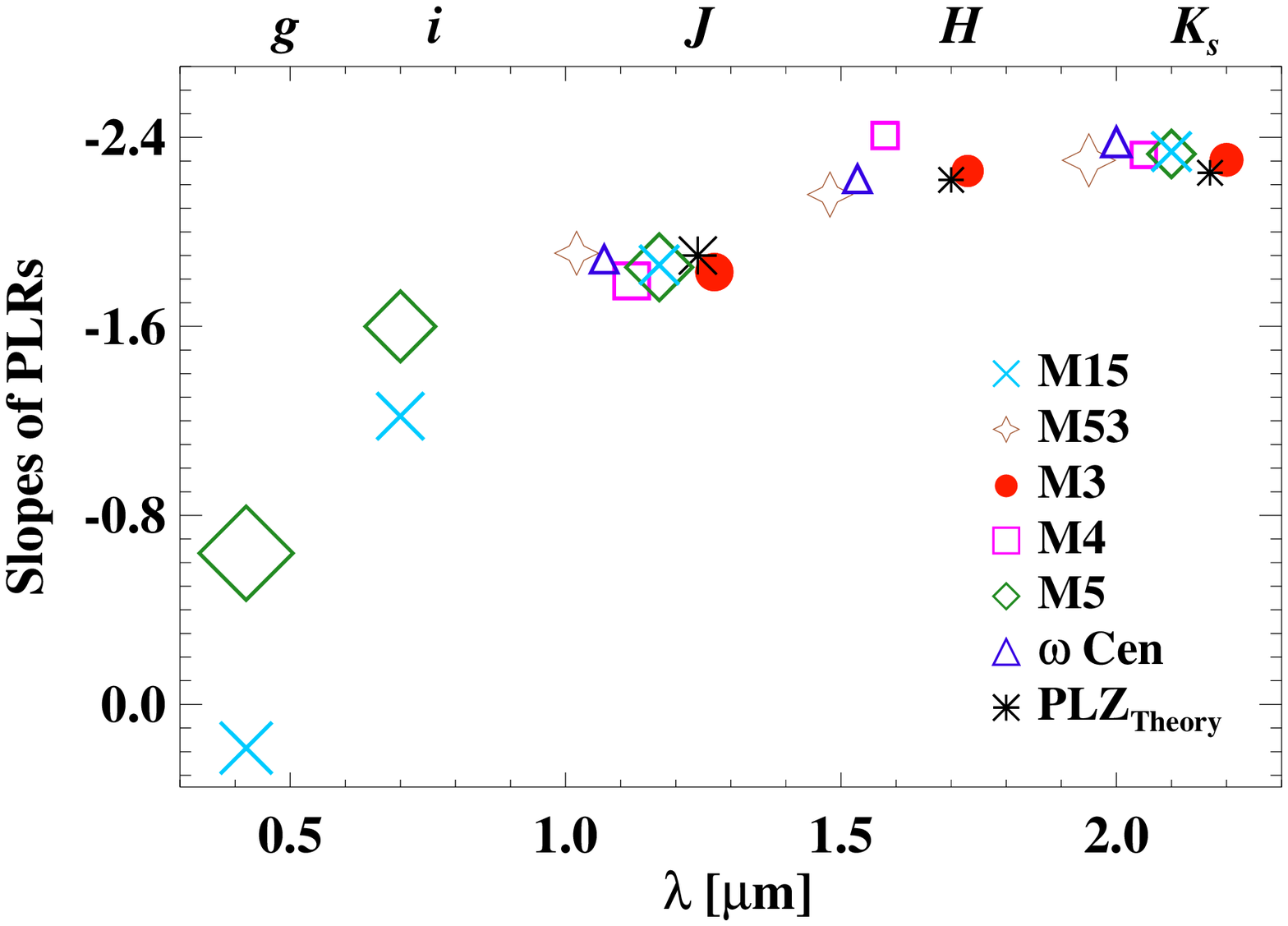}
\caption{A comparison of the slopes of $gi$ and $JHK_s$ PLRs of RR Lyrae variables in different GCs. The data in the NIR filters are slightly shifted along the {\it x}-axis for visual clarity.  The symbol size increases as a function of uncertainty in the slope of the PLR in a given GC. The empirical slopes are from the literature as follows:  M53 \citep{bhardwaj2021}, M5 \citep{coppola2011, vivas2017}, M4 \citep{braga2015}, $\omega$ Cen \citep{braga2018}, and M3 \citep{bhardwaj2020a}. The theoretical PLZ slopes are adopted from \citet{marconi2015}.}
\label{fig:slp_plr}
\end{figure}

Figure~\ref{fig:slp_plr} shows the slopes of the global sample of RR Lyrae stars in different GCs. The difference in the slopes of the PLRs for RR Lyrae stars in M5 and M15 decreases moving from optical to NIR bands. The larger difference in the slopes at shorter wavelengths could be due to the increased sensitivity to the metallicity difference between OoI and OoII clusters. However, the slopes of optical band PLRs also exhibit larger uncertainties and are statistically similar in the $i$-band. The slopes reach an asymptotic value of $\sim-2.3$ in the NIR, which confirms the ﬂattening of the period dependence at longer wavelengths \citep{neeley2017}. The slopes of $JHK_s$-band PLRs are statistically consistent given their uncertainties, and no trend is seen as a function of metallicity or Oosterhoff type of the cluster.

\begin{figure}
\epsscale{1.2}
\plotone{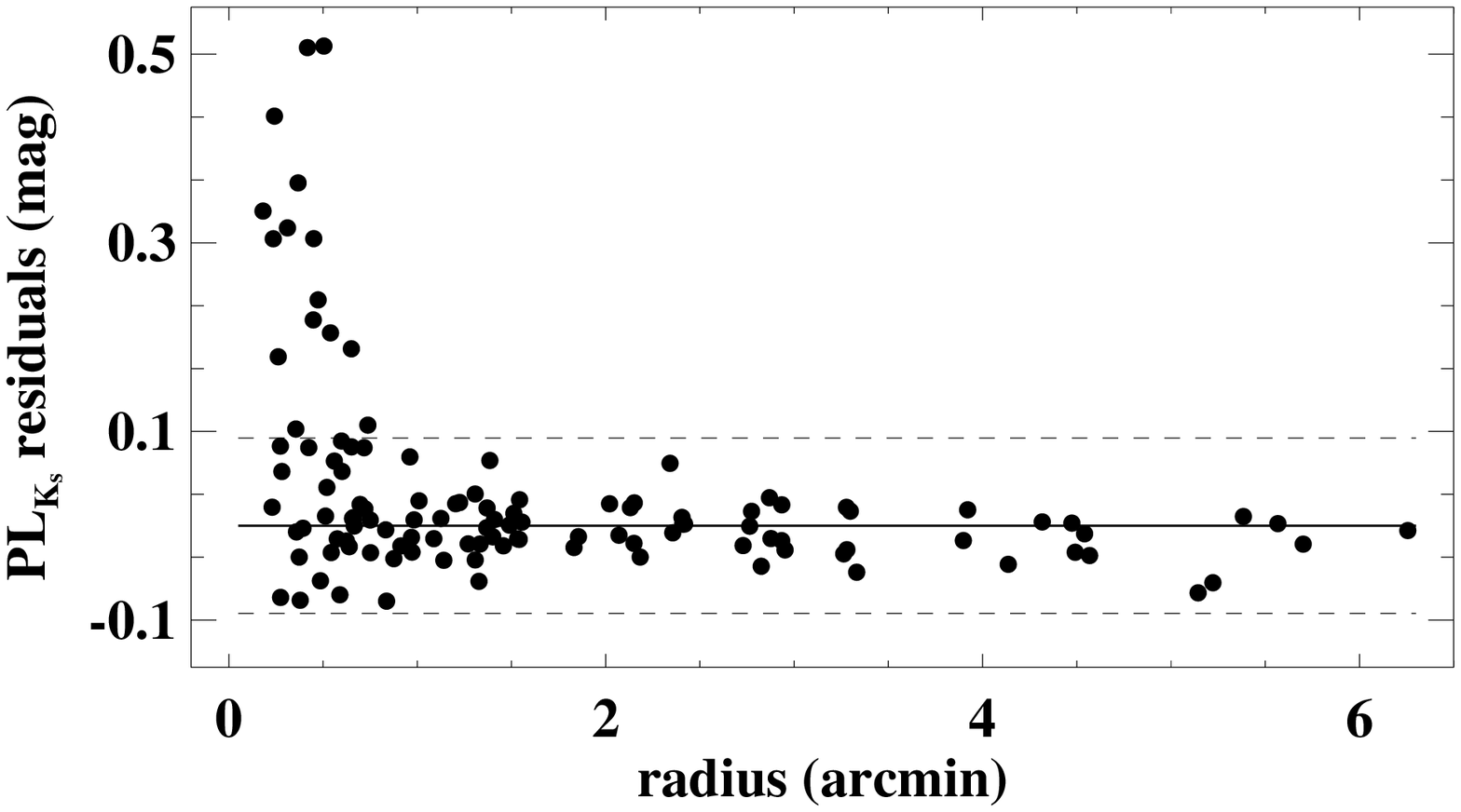}
\plotone{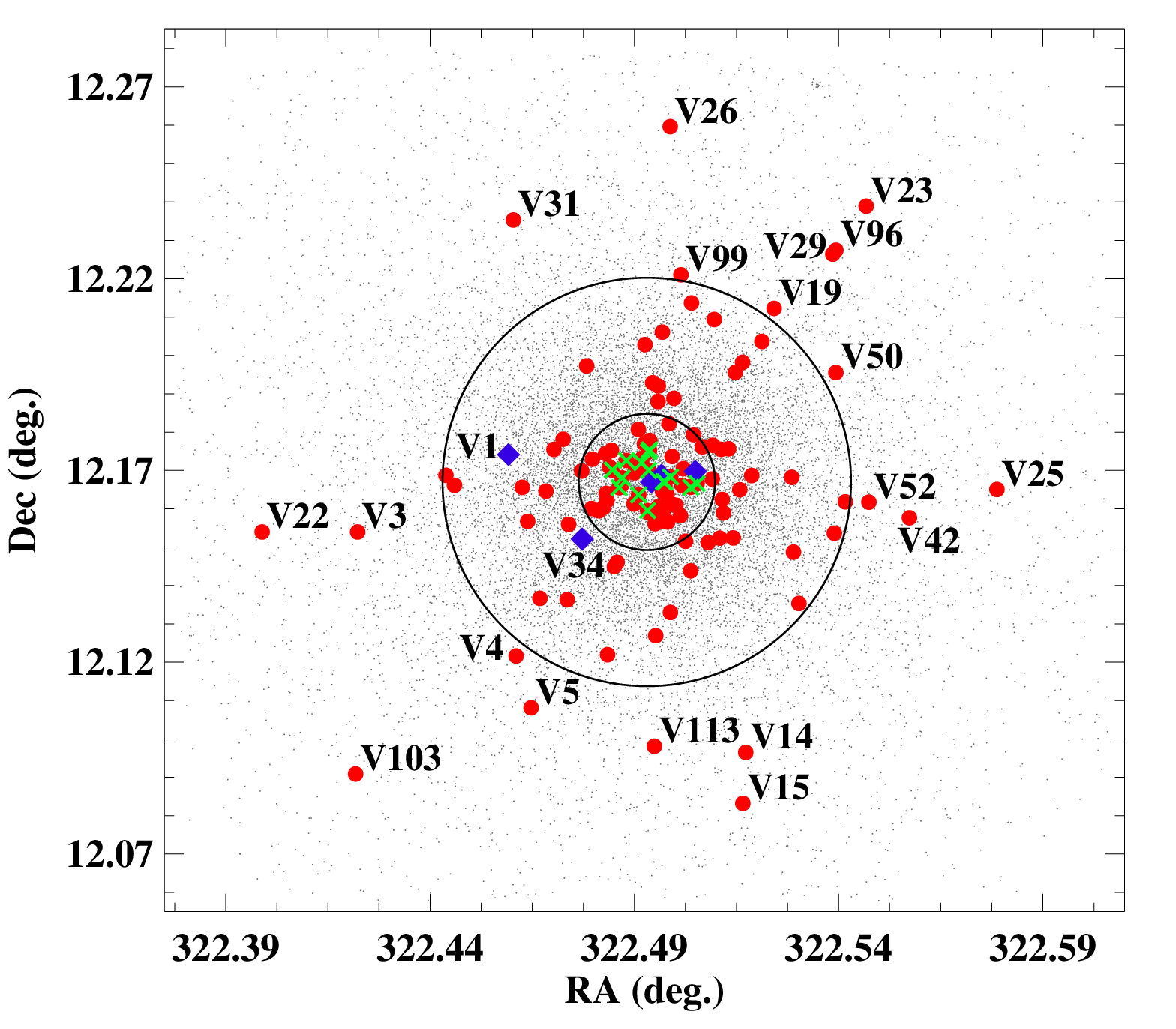}
\caption{{\it Top:} Residuals of the $K_s$-band PLR as a function of radial distance from the cluster center. The solid line represents a residual of zero, while dashed lines show $\pm2.5\sigma$ scatter in the residuals. {\it Bottom:} The spatial distribution of all sources in M15 within the FoV of WIRCam (grey dots) with RR Lyrae stars (red circles), and Cepheid variables (blue squares). The outliers (green crosses) from the best-fitting $J$ and/or $K_s$-band PLRs are also shown. The small and big circles represent the half-light radius \citep[$r_h=1\arcmin$,][]{harris2010} and the $3r_h$ radius, respectively. RR Lyrae stars outside $3r_h$ and Cepheid stars outside $1r_h$ are labeled.} 
\label{fig:spatial}
\end{figure}

\begin{figure}
\epsscale{1.2}
\plotone{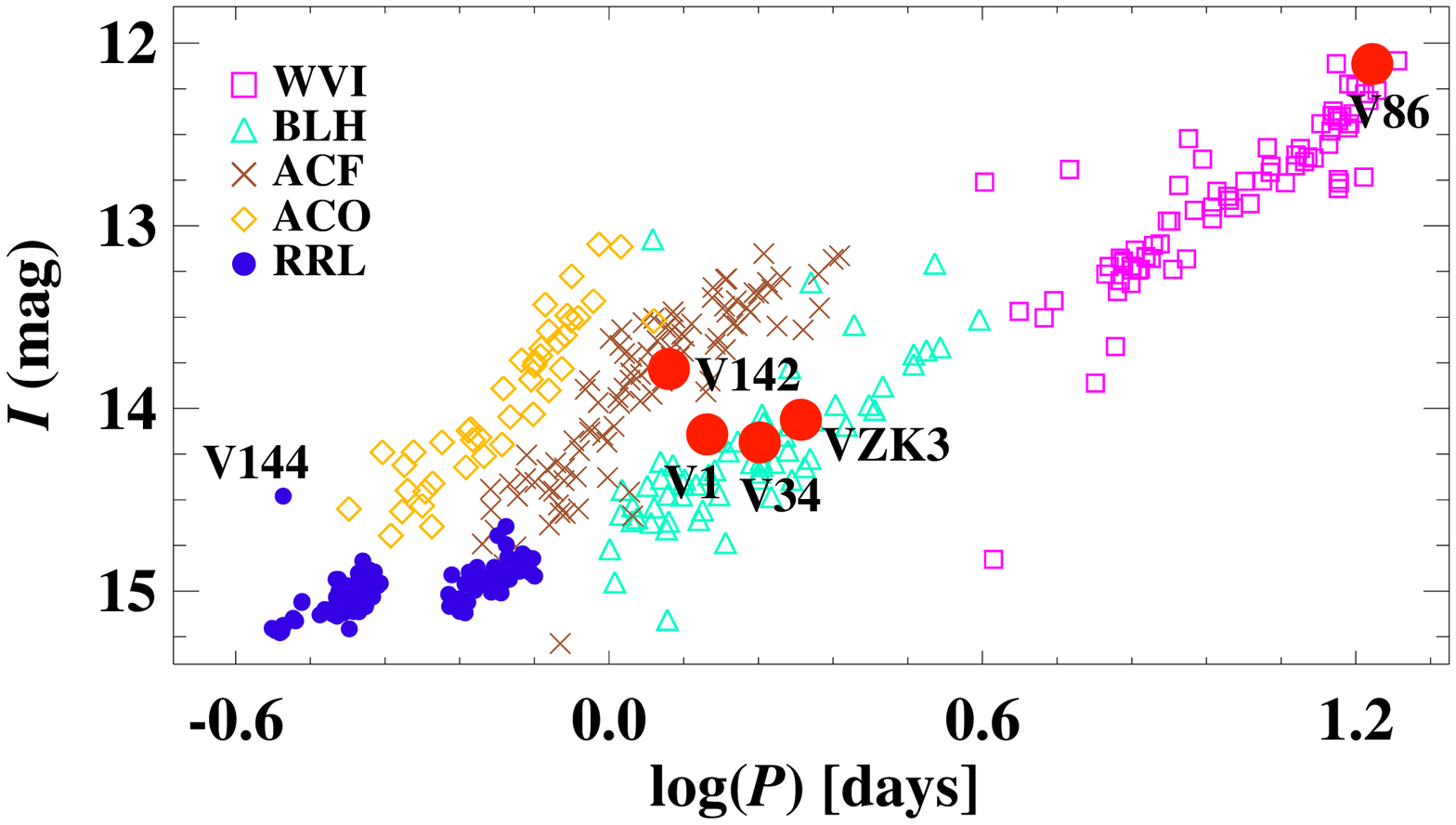}
\plotone{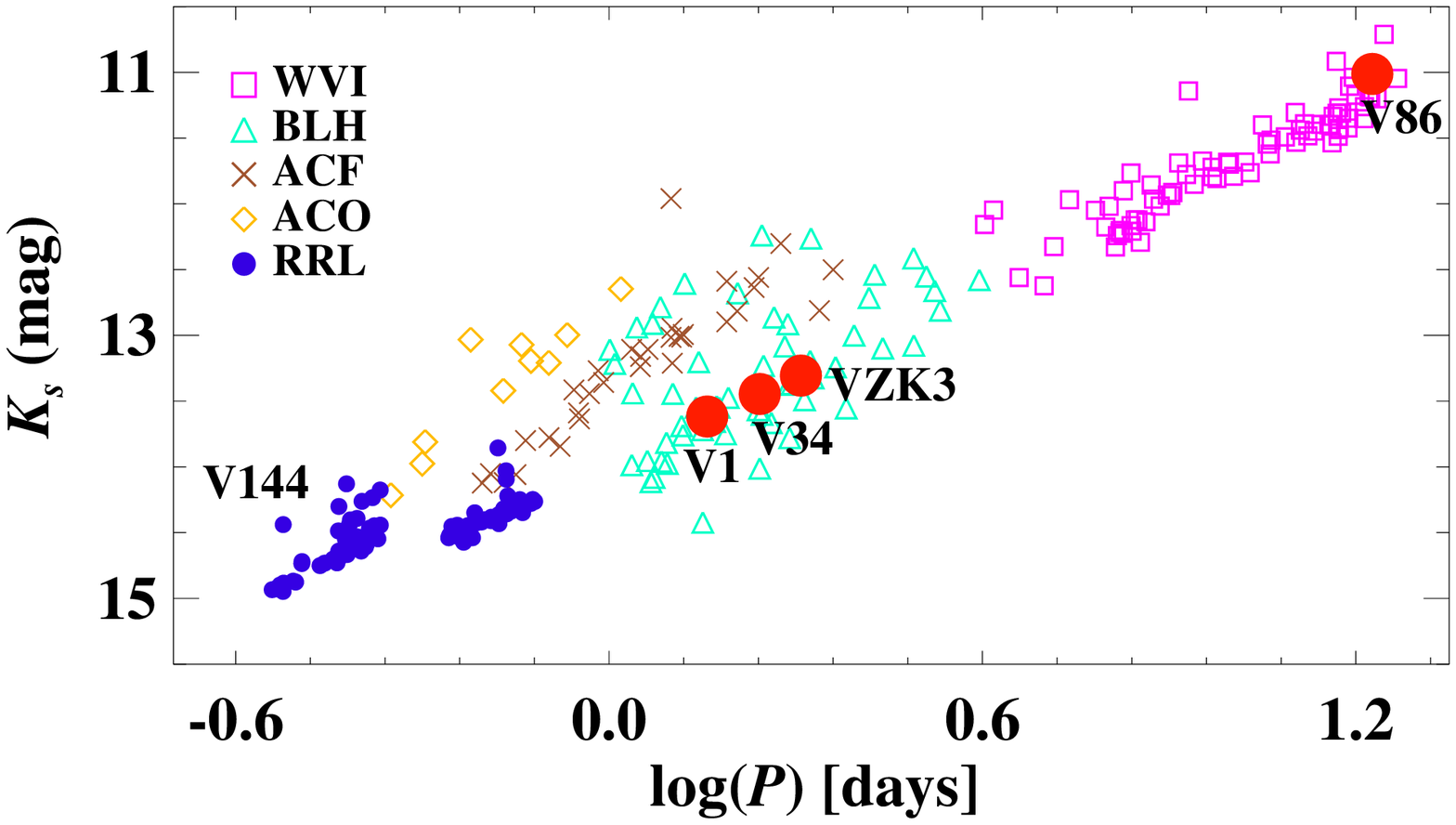}
\caption{Period-luminosity relations for Population II Cepheids and RR Lyrae variables in $I$ (top) and $K_s$ (bottom). The $I$-band mean magnitudes for BL Herculis (BLH), W Virginis (WVI), and the fundamental and first-overtone mode anomalous Cepheids (ACF and ACO) are from the LMC \citep{soszynski2015a, soszynski2018}. RR Lyrae stars and Cepheid candidates (large red circles) in M15 are also shown. Note that SDSS $i$-band magnitudes for M15 variables are transformed to the $I$ filter for a relative comparison. The mean magnitudes at $K_s$ for Type II Cepheids and anomalous Cepheids are from \citet{bhardwaj2017} and \citet{ripepi2014a}, respectively.}
\label{fig:plr_ik}
\end{figure}

The minimal scatter ($\sigma \leq 0.04$~mag) in NIR PLRs is expected since the mean magnitudes are well determined and there is no significant spread in metallicity of M15. \citet{carretta2009} found a mean value of [Fe/H]$=-2.33\pm0.02$~dex ($\sigma=0.06$~dex) based on intermediate-to-high resolution spectra of 84 red giants in M15. However, several RR Lyrae stars are found to be significantly brighter than the best-fitting PLRs in Figure~\ref{fig:jk_plr}. We suspect that these stars are likely blended sources because all of these exhibit good-quality light curves. To investigate the impact of crowding in the central region, we plotted the residuals of the $K_s$-band PLR as a function of their radial distance from the center of the cluster in the top panel of Figure~\ref{fig:spatial}. It becomes clear that all $2.5\sigma$ outliers in the $K_s$-band PLR are located within $\sim 0.7\arcmin$ radius from the cluster center. No trend is seen in the residuals of RR Lyrae stars located in the outskirts of the cluster.  We do not find any statistically significant difference in the slope or zero-point of the PLRs in the Table~\ref{tbl:plrs} when excluding the sources within $\sim 0.7\arcmin$ radius from the center of the cluster. The bottom panel of Figure~\ref{fig:spatial} shows the spatial distribution of all M15 sources within the FoV of WIRCam. Note that three of our candidate Cepheid variables are also located within the crowded half-light radius of the cluster.

We also investigated the PLRs of candidate Cepheid variables in Figure~\ref{fig:plr_ik}. The mean magnitudes at $I$ for Cepheids in the LMC were taken from the OGLE survey \citep{soszynski2015a, soszynski2018} and have been corrected for a relative distance modulus between the LMC \citep[$\mu=18.477\pm0.026$,][]{piet2019} and M15 \citep[$\mu=15.15\pm 0.02$,][]{baumgardt2021} of $\sim3.33$ mag. Furthermore, SDSS $i$-band magnitudes were transformed to $I$ using transformations from \citet{jordi2006}. For this purpose a mean value of $(i-z)=0.06$ mag was adopted \citep[$-0.15<(i-z)<0.15$ mag;][]{an2008}, which can lead to a maximum systematic uncertainty of 0.06~mag in the transformed magnitudes. After applying extinction corrections, we find that V86 falls perfectly on the W Virginis PLR. Three other Cepheid candidates (V1, V34, and VZK3) are consistent with the $I$-band PLR for BL Herculis variables in the LMC. However, V142 is relatively brighter in the $I$ band as compared to other short-period ($P\sim1-2$~days) Cepheid candidates. We classify it as a fundamental-mode anomalous Cepheid based on its location on the $I$-band PLR, although it falls in the overlapping region for anomalous Cepheid and BL Herculis stars in the period--amplitude diagram. The $gi$ light curves of V142 are nearly identical to V1. Furthermore, it is also located within the central $1\arcmin$ (see Figure~\ref{fig:spatial}) and may be blended with a bright source. In that case, V142 could be a BL Herculis variable as classified in the catalog of \citet{clement2001}.

The bottom panel of Figure~\ref{fig:plr_ik} shows the $K_s$-band PLR for known Type II and anomalous Cepheids in the LMC from \citet{bhardwaj2017} and \citet{ripepi2014a}, respectively. While data from \citet{bhardwaj2017} is in the 2MASS system, VMC survey photometry \citep{ripepi2014a} is in the VISTA system. The latter was also converted into the 2MASS system using the empirical transformations{\footnote{\url{http://casu.ast.cam.ac.uk/surveys-projects/vista/technical/photometric-properties}}}. Once the relative distance and the extinction corrections are applied, all Cepheid candidates observed in the NIR data fall right on the Type II Cepheid PLR. Considering the optical and NIR periodic light curves (Figs.~\ref{fig:lcs_cep_gi} and \ref{fig:lcs_cep_jk}) and consistency in the amplitudes and mean magnitudes, we classify these variables as Type II Cepheids. However, we again emphasize that V86, V142, and VZK3 are located in the unresolved central $1\arcmin$ (see Figure~\ref{fig:spatial}), and multi-epoch photometry with higher angular resolution imaging would help to properly resolve and classify these candidate Cepheid variables.

\section{Distance to M15}
\label{sec:dis}

M15 is a well-studied GC which has several distance determinations from independent methods in the literature that vary typically between 10 and 11.5 kpc. \citet{harris2010} catalogued a distance of 10.4 kpc to M15 which is the most commonly adopted value in the literature. Table~\ref{tbl:mu_m15} lists a few recent determinations of distance modulus to M15 based on different methods. Most of the RR Lyrae distances range between $15.1$ and $15.2$ mag. \citet{sollima2006} derived a distance modulus of 15.13 mag using $K_s$-band observations of RR Lyrae stars in GCs of different mean metallicities. \citet{dambis2014} found that the distance modulus varies by 0.32 mag depending on the adopted absolute calibration of mid-infrared PLRs of RR Lyrae stars. The calibration of RR Lyrae PLRs based on the parallaxes from the {\it Hubble Space Telescope {(HST)}}  results in statistically consistent distance moduli by \citet{dambis2014} and \citet{benedict2011}. Recently, \citet{baumgardt2021} used {\it Gaia} EDR3, {\it HST} kinematics data, and several distance moduli from the literature to obtain a mean value of $15.15\pm0.02$~mag to M15. We use our optical and NIR data to derive a distance to M15 in the following sections.

\begin{deluxetable}{CClc}
\tablecaption{True distance modulus to the M15 cluster. \label{tbl:mu_m15}}
\tabletypesize{\footnotesize}
\tablewidth{0pt}
\tablehead{\colhead{$\mu$} & \colhead{[Fe/H]} & \colhead{Method} & \colhead{Ref.}\\
	mag		&	 		&		 	&	\\}
\startdata
15.15$\pm$0.02  &   -       &   {\it Gaia} EDR3 + {\it HST} + $\mu_{\rm literature}$  &B21    \\ 
15.26$\pm$0.03	&	-2.36	&	{\it HST} CMD \& Isochrone fitting		&V20	\\
15.22$\pm$0.05  &   -       &   RRL density model fitting           &H19   \\ 
15.16$\pm$0.01	&	-2.36	&	{\it HST} CMD \& Isochrone fitting		&W17	\\
15.38$\pm$0.10	&	-2.30	&	Main-sequence fitting		        &O17$^a$	\\
15.25$\pm$0.06	&	 -	    &	Type II Cepheid PLR$_{K_s}$ 		            &B17	\\
15.11$\pm$-	    &	-2.30	&	{\it HST} CMD \& Isochrone fitting		&V16	\\
15.14$\pm$0.03	&	-2.37	&	MIR RRL PLRs + $HST_\pi$		    &D14	\\
14.82$\pm$0.03	&	-2.37	&	MIR RRL PLRs 		                &D14	\\
15.19$\pm$0.11	&	-2.16	&	RR Lyrae PLZ$_{K_s}$ relation       &B11$^b$	\\
14.99$\pm$0.11  &	-2.42	&	M$_V-$[Fe/H]  relation              &B07	\\
15.13$\pm$0.13	&	-2.15	&	RR Lyrae PLZ$_{K_s}$  relation      &S06$^b$	\\
15.13$\pm$-     &	-2.26	&	M$_V-$[Fe/H]  relation 		                &M06	\\
14.79$\pm$0.10	&	-1.92	&	Fourier decomposition 		        &AF06	\\
15.11$\pm$0.06  &	-2.33	&	RR Lyrae PLZ$_i$  relation      &TW	\\
15.22$\pm$0.05  &	-2.33	&	RRc PLZ$_J$  relation             &TW	\\
15.19$\pm$0.03  &	-2.33	&	RRab PLZ$_J$  relation              &TW	\\
15.20$\pm$0.03  &	-2.33	&	RR Lyrae PLZ$_J$  relation         &TW	\\
15.20$\pm$0.04  &	-2.33	&	RRc PLZ$_{K_s}$  relation           &TW	\\
15.19$\pm$0.03  &	-2.33	&	RRab PLZ$_{K_s}$  relation         &TW	\\
15.20$\pm$0.03  &	-2.33	&	RR Lyrae PLZ$_{K_s}$  relation    &TW	\\
15.18$\pm$0.03  &	-2.33	&	RR Lyrae PLZ$_{K_s}$ + {\it Gaia} EDR3 &TW	\\
15.13$\pm$0.05  &	-	    &	Type II Cepheid PLR$_{J}$  relation &TW	\\
15.18$\pm$0.04  &	-	    &	Type II Cepheid PLR$_{K_s}$  relation    &TW	\\
\hline
\multicolumn{3}{l}{$\mu_\textrm{M15} = 15.196~\pm~0.026$~(stat.)~$\pm~0.039$~(syst.)~mag}       &TW\\
\hline
\multicolumn{3}{l}{$D_\textrm{M15} = 10.944~\pm~0.131$~(stat.)~$\pm~0.187$~(syst.)~kpc} &TW\\
\enddata
\tablecomments{Most of these studies adopted E(B-V)=0.1~mag \citep{harris2010} except where specified with the following notes - $^a$E(B-V)=0.11 mag, $^b$E(B-V)=0.09 mag. References - B21 - \citet{baumgardt2021}, V20 - \citet{valcin2020}, H19 -\citet{hernitschek2019}, W17 - \citet{wagner2017}, O17 - \citet{omalley2017},  B17 - \citet{bhardwaj2017}, V16 - \citet{vandenberg2016}, D14 - \citet{dambis2014}, B11 - \citet{benedict2011}, B07 - \citet{bono2007}, S06 - \citet{sollima2006},  M06 - \citet{matsunaga2006}, AF06 - \citet{ferro2006}, and TW - this work.} 
\end{deluxetable}

\subsection{RR Lyrae-based distance using NIR period-luminosity-metallicity relations} 

The new NIR photometry for RR Lyrae stars in M15 provides an opportunity to determine a robust distance. Previous distance measurements based on NIR PLRs for RR Lyrae stars used a sample of 52 stars with single-epoch observations from \citet{sollima2006}. The $K_s$-band photometry for 20 stars by \citet{longmore1990} was utilized together with a small sample of 5 RR Lyrae with {\it HST} parallaxes from \citet{benedict2011}. While these studies used an empirical calibration of the PLZ$_{K_s}$ relation for RR Lyrae stars, most recent studies on distance scale have used theoretical calibrations \citep{marconi2015} which predict a relatively larger metallicity coefficient for the PLZ$_{K_s}$ relation than the empirical relations \citep[e.g.,][]{muraveva2015, navarrete2017, braga2018}. Recently, \citet{bhardwaj2021} used multi-epoch $K_s$-band photometry in 5 GCs to derive an empirical PLZ$_{K_s}$ relation for RR Lyrae stars where metallicity dependence is consistent with theoretical predictions.

We used theoretical PLZ relations for RR Lyrae stars at $J$ and $K_s$ from \citet{marconi2015} to derive distance moduli to M15 using different samples of RRab, RRc, and all RR Lyrae stars. The mean metallicity of [Fe/H]$=-2.33\pm0.02$~dex \citep{carretta2009} was used for all RR Lyrae stars in M15. Absolute magnitudes were determined for each RR Lyrae variable using theoretical $J$- and $K_s$-band PLZ relations given their pulsation periods and extinction-corrected mean magnitudes. A weighted mean distance modulus was obtained by excluding the RR Lyrae stars that are outliers in the PLRs. Table~\ref{tbl:mu_m15} lists the distance moduli for each sample. These values are statistically similar within $0.5\sigma$ of their quoted uncertainties and are consistent with other measurements based on different methods.

The uncertainties in the distance moduli based on accurate and precise NIR data for RR Lyrae stars are limited mostly to the absolute zero-point calibration of their PLRs. Therefore, we also employ an empirical calibration of the PLZ$_{K_s}$ relation for RR Lyrae variables from \citet{bhardwaj2021} based on GC data and {\it Gaia} EDR3 data. We find a distance modulus of $15.18\pm0.03$~mag, in excellent agreement with theoretical calibrations. A true distance modulus of $15.196~\pm~0.026~{\rm(stat.)}~\pm~0.039~{\rm(syst.)}$~mag to M15 was obtained by taking a weighted mean of all RR Lyrae-based NIR measurements. The systematic uncertainties were obtained by adding in quadrature the errors in the zero-points, errors in the slope of the calibrator and M15 PLRs, and uncertainties due to variations in the mean metallicity and extinction. Our distance to M15 of $10.944~\pm~0.131~{\rm(stat.)}~\pm~0.187~{\rm(syst.)}$~kpc is consistent with the mean distance to M15 derived by \citet{baumgardt2021} based on several distance moduli in the literature.

\subsection{Population II Cepheid-based distance using NIR period-luminosity relations} 

NIR photometry for Type II Cepheids in M15 can also be used to determine an accurate distance modulus. We used the empirical $JK_s$-band PLRs for the combined sample of BL Herculis and W Virginis stars in the LMC as calibrators from \citet{bhardwaj2017} anchored with the distance from late-type eclipsing binaries \citep{piet2019}. \citet{matsunaga2006} found no metallicity dependence on NIR Type II Cepheid PLRs in GCs and thus no metallicity term is included in the calibrator PLRs. Furthermore, \citet{bhardwaj2017b} showed that the slopes and zero-points of the calibrated $K_s$-band PLR for the combined sample of BL Herculis and W Virginis stars 
are statistically consistent in the Galactic bulge, Galactic GCs, and the LMC.

We used calibrated PLRs from the LMC to determine an absolute magnitude for all four Type II Cepheids in M15. A weighted mean  is taken in the $J$ and $K_s$ bands separately, and Table~\ref{tbl:mu_m15} lists the resulting distance moduli values. The Cepheid-based distance moduli agree with most independent distances listed in Table~\ref{tbl:mu_m15}, although the uncertainties are relatively larger given the small statistics. \citet{bhardwaj2017} derived a distance of $15.25\pm0.06$ using NIR photometry for the W Virginis star V86 in M15 from \citet{matsunaga2006}. Our distance modulus based on the $K_s$-band Type II Cepheid PLR agrees very well with our adopted RR Lyrae distance to M15.

\subsection{RR Lyrae-based distance using $i$-band period-luminosity-metallicity relations} 

We also utilized optical photometry for RR Lyrae stars to derive a distance to M15. \citet{caceres2008} presented theoretical PLZ relations for RR Lyrae variables in the SDSS photometric systems which we employ for the calibration of the PLRs derived in Section~\ref{sec:plr}. The slope of the empirical $i$-band PLR is statistically consistent with the theoretical PLZ relation of \citet{caceres2008}. To derive absolute magnitudes for RR Lyrae stars in the $i$ band, we used the mean metallicity  [Fe/H]$\sim-2.33\pm0.02$~dex \citep{carretta2009} for M15 to measure $\log Z$ using equations (4) and (5) of \citet{caceres2008}. For this purpose, we adopted an enhancement of $\alpha$ elements, $[\alpha/{\rm Fe}] = 0.4$ \citep{pritzl2005, vandenberg2016}.

The extinction-corrected mean magnitudes at $i$ for RR Lyrae  stars were used to obtain a distance modulus of $15.11\pm0.06$~mag to M15, which falls within the typical range of distance moduli values in Table~\ref{tbl:mu_m15}.
\citet{caceres2008} suggested that their predicted PLZ relation at $i$ is an average relation that is less precise than the theoretical relations involving a color term using bluer SDSS colors. Furthermore, the resulting distance modulus also increases if a smaller value for $[\alpha/{\rm Fe}]$ is used to estimate $\log Z$, and \citet{valcin2020} determined a $[\alpha/{\rm Fe}]$ value as small as 0.18~dex for M15. Therefore, no proper account of systematic uncertainties associated with the calibration of $i$-band PLZ relations is provided, and the distance modulus is statistically consistent with our adopted RR Lyrae-based distance to M15.

\section{Summary} \label{sec:discuss}

We reported new optical and NIR multi-epoch observations for variable stars in the globular cluster M15 using data from the CFHT science archive. The variables in our sample include 129 RR Lyrae, 3 BL Herculis, 1 W Virginis, and 1 anomalous Cepheid candidate. This sample is the largest with NIR multi-epoch photometry for variable stars in this cluster. Since our photometric data is obtained with a 3.6m class telescope and covers a long temporal baseline, periods and classification of several variable sources are improved, particularly in the inner region, despite the limited number of epochs in the dataset.

Optical and NIR photometry were used to study pulsation properties of RR Lyrae and Cepheid variables. The horizontal branch of M15 spans a wide color range, and is well-populated with variable RR Lyrae stars. Cepheid candidates are significantly brighter than the horizontal-branch stars in both the optical and NIR color--magnitude diagrams. The location of RR Lyrae variables in both the optical and NIR color--magnitude diagrams aligns closely with the predicted boundaries of the instability strip. The Bailey diagrams of RR Lyrae stars in M15 show that most of these variables fall on the locus of OoII RRab stars similar to other Oosterhoff Type II GCs. We did not find any variation in the optical-to-NIR amplitude ratios for RRab stars in M15 as a function of pulsation periods. However, these amplitude ratios are smaller for RRc stars than for the RRab stars as observed for RR Lyrae stars in M3 and $\omega$ Cen.

New NIR time-series data for RR Lyrae stars were used to derive precise PLRs in a metal-poor GC. In optical bands, a tight PLR was observed only in the $i$ band, and the slope of the $g$-band PLR is statistically consistent with zero, implying little or no dependence on pulsation period. The PLR at $K_s$ exhibits a scatter of only $0.037$ mag, which reduces to only $0.02$~mag if a stricter outlier removal threshold is adopted. Using RR Lyrae PLRs, we determined a true distance modulus to M15 of $15.196~\pm~0.026~{\rm(stat.)}~\pm~0.039~{\rm(syst.)}$~mag. Our distance to M15 based on RR Lyrae and Type II Cepheid variables agrees well with most distance measurements for M15 available in the literature.

Our NIR photometry for RR Lyrae stars will be important for constraining the metallicity dependence of empirical PLZ relations at the metal-poor end of the RR Lyrae metallicity distribution. Unlike other metal poor Galactic GCs, M15 has a rich RR Lyrae population and will significantly increase the statistics of metal-poor RR Lyrae stars when combined with similar data in other GCs with NIR multi-epoch observations \citep{bhardwaj2021}. The absolute calibration of RR Lyrae PLZ relations at NIR wavelengths is essential for their application as primary calibrators for the first-rung of the Population II distance ladder. An accurate and precise distance ladder calibrated with Population II standard candles will offer a unique opportunity to test the consistency of the traditionally adopted classical Cepheid-based route to the Hubble constant.

\acknowledgements

We thank the anonymous referee for the constructive and useful report that helped improve the manuscript. AB acknowledges a Gruber fellowship 2020 grant sponsored by the Gruber Foundation and the International Astronomical Union and is supported by the EACOA Fellowship Program under the umbrella of the East Asia Core Observatories Association, which consists of the Academia Sinica Institute of Astronomy and Astrophysics, the National Astronomical Observatory of Japan, the Korea Astronomy and Space Science Institute, and the National Astronomical Observatories of the Chinese Academy of Sciences. This research was supported by the Munich Institute for Astro- and Particle Physics (MIAPP) of the DFG cluster of excellence ``Origin and Structure of the Universe''. 

This work has made use of data from the European Space Agency (ESA) mission {\it Gaia} (\url{https://www.cosmos.esa.int/gaia}), processed by the {\it Gaia} Data Processing and Analysis Consortium (DPAC, \url{https://www.cosmos.esa.int/web/gaia/dpac/consortium}). Funding for the DPAC has been provided by national institutions, in particular the institutions participating in the {\it Gaia} Multilateral Agreement. 

\facility{CFHT (MegaCam and WIRCam imagers)}

\software{\texttt{IRAF} \citep{tody1986, tody1993}, \texttt{DAOPHOT/ALLSTAR} \citep{stetson1987} \texttt{DAOMATCH and DAOMASTER} \citep{stetson1993}, \texttt{ALLFRAME} \citep{stetson1994}, \texttt{SExtractor} \citep{bertin1996}, \texttt{SWARP} \citep{bertin2002}, \texttt{SCAMP} \citep{bertin2006}, \texttt{WeightWatcher} \citep{marmo2008}, \texttt{IDL} \citep{landsman1993}, \texttt{Astropy} \citep{astropy2013}}\\\\\\\\

\appendix

\section{Comments on individual variable stars}
\label{sec:comments}

In this section, we provide some specific comments on a few variable stars from the list of \citet{clement2001}. Out of 191 variable candidates listed in their catalogue, we presented periods and classifications for 134 RR Lyrae and Population II Cepheid variables. Only two variables (V104 and V105) are outside of both the MegaCam or WIRCam fields of view. Other remaining known candidate variable stars cross-matched with the catalog of \citet{clement2001} are not discussed in this work for one and more of the following reasons: (1) light curves exhibit scatter with large uncertainties on individual photometric measurements; (2) no evidence of periodicity is found; (3) variables fall within the gaps between the detectors or towards the edges of the detectors and are therefore observed in less than half of the total number of frames; (4) blending in the crowded central regions has made the photometry unreliable; (5) they are located outside the FoV or no suitable match is found within $2\arcsec$ based on magnitude and color cuts for candidate variables. 

{\it V1}---A known short-period Type II Cepheid  \citep{clement2001, siegel2015, hoffman2021}. We confirm the classification of V1 as a BL Herculis star, a subclass of Type II Cepheids, based on the optical and NIR PLRs.

{\it V34}---A suspected eclipsing binary star  \citep{clement2001} with a period of 1.1591 days. \citet{siegel2015} find a period of 2.037 days and suggested that it is an anomalous Cepheid. \citet{hoffman2021} determined a period 0.40096 days and did not provide a classification. We classify it as a BL Herculis subclass of Type II Cepheid with a period of 2.03355 days based on its location on the color--magnitude diagrams, peak-to-peak amplitudes and mean magnitudes on the PLRs.

{\it V72}---It has a period of 1.1386 days  \citep{clement2001}, but no classification is provided. \citet{hoffman2021} classified V72 as RRab, but their light curve exhibits large scatter and does not show any evidence of a typical saw-tooth structure. We find a period of 0.39549 days and classify V72 as an RRc star. While the phased light curve is nearly sinusoidal with relatively larger scatter, its mean magnitudes fall on the NIR PLR for RR Lyrae stars for our adopted period.

{\it V86}---A known long-period Type II Cepheid  \citep{clement2001, siegel2015, hoffman2021}. We confirm classification of V86 as a W Virginis star, a subclass of Type II Cepheids, based on the peak-to-peak amplitudes and the optical and NIR PLRs.

{\it V142}---It is classified as a Type II Cepheid \citep{clement2001}. It was not observed by \citet{siegel2015} and \citet{hoffman2021}. While it was not observed with our NIR data, optical light curves show a periodicity of 1.24805 days. Based on the $i$-band PLR, it is tentatively classified as a fundamental-mode anomalous Cepheid. However, it is located within the central $1\arcmin$ radius of the cluster and could be brighter than the Type II Cepheid PLR due to a blend. In that case, V142 could be a BL Herculis variable.

{\it V144}---An RRc star with clearly periodic light curves with a period of 0.29949 days, slightly different to the one listed by \citet{clement2001}. This star was not observed by \citet{siegel2015} or \citet{hoffman2021}. V144 is brighter than most RRc stars in both the optical and NIR color--magnitude diagrams, and is located in the crowded central region. We suspect its photometry is blended with a bright source, but it could also be a foreground RRc variable.

{\it V155}---\citet{clement2001} did not report a period for this variable and \citet{siegel2015} did not include it. \citet{hoffman2021} suggested that it is a Type II Cepheid with period of 0.91189 days. We find clear saw-tooth light curves and classify it as an RRab star with a period of 0.61251 days.

{\it VZK3}---It is classified as an RRab star by  \citet{clement2001}. We find it to be brighter than the horizontal-branch stars and classify it as a BL Herculis star with a period of 1.74634 days. However, it is located in the central region of the globular cluster and could be brighter due to blending with a nearby source. In that case, VZK3 could also be an RRc star because its periodogram shows a secondary peak at 0.40527 days and the light curves are almost sinusoidal exhibiting amplitudes consistent with the RRc stars.

{\it V182}---\citet{siegel2015} found a new RRc star with a period of 0.39139 days ($\alpha=$ 21:30:02.93, $\delta=$+12:10:09.5). However, we suspect that it is the same as VN9 ($\alpha=$ 21:30:02.87, $\delta=$+12:10:08.8, Period = 0.3915 days) from the catalog of \citet{clement2001} considering its period and coordinates ($\Delta = 1.1\arcsec$). 

{\it V15, V17, V21, V32, V37, V40, V41, V44, V48, V53, V57, V60, V67, V71, V73, V82, V88, V89, V100, V107, V116, V120, V129, V130, V163, V166, V168, V173, V177, VZK4, VZK74, VZK78, VNV3, VNV11}---The phased light curves of these variables exhibit significantly smaller scatter when phased with our adopted periods listed in Table~\ref{tbl:m15_all} as compared to those from \citet{clement2001}.

{\it V27, V79, V85, V95, V143, V146, V147, V148, V149, V150, V151, V153, V154}---We confirm that these are non-variable stars as mentioned in the catalog by \citet{clement2001} catalog.

{\it V98, V108, V110, V111, V114, V117, V119, V176, V182}---Our photometric light curves of these stars are of good quality, but we do not find any evidence of variability and periodicity. \citet{hoffman2021} found periodicity for a few of these stars, but their light curves exhibit large scatter. These RR Lyrae candidates are located in the unresolved central region, where our photometric data may be limited by crowding. Higher resolution time-series photometry will be useful to confirm the variable nature of all the sources that are not discussed for the reasons outlined above in the Appendix~\ref{sec:comments}.

\bibliographystyle{aasjournal}
\bibliography{mybib_final.bib}
\end{document}